\documentclass[fleqn,usenatbib]{mnras}

\usepackage{mathptmx}

\usepackage[T1]{fontenc}

\DeclareRobustCommand{\VAN}[3]{#2}
\let\VANthebibliography\thebibliography
\def\thebibliography{\DeclareRobustCommand{\VAN}[3]{##3}\VANthebibliography}

\usepackage{graphicx}	
\usepackage{amsmath}	
\let\Bbbk\relax
\usepackage{amssymb}	

\title[G3 PINOCCHIO]{Numerical implementation of the Cubic Galileon model in {\sc pinocchio}}

\author[Y. Song et al.]{
Yanling Song,$^{1}$
Chiara Moretti,$^{2,4}$
Pierluigi Monaco$^{3,4,5,6}$
and Bin Hu$^{1}$\thanks{E-mail: bhu@bnu.edu.cn}
\\
$^{1}$Department of Astronomy, Beijing Normal University, Beijing 100875, China\\
$^{2}$ Astronomy Unit, School of Physics and Astronomy, Queen Mary University of London,\\
Mile End Road, London, E1 4NS, UK
$^{3}$Dipartimento di Fisica dell'Universit\'a di Trieste, Sezione di
Astronomia, via Tiepolo 11, I-34143 Trieste, Italy\\ 
$^{4}$INAF -- Osservatorio Astronomico di Trieste, Via Tiepolo 11, I-34143 - Trieste,
Italy\\ 
$^{5}$IFPU -- Institute for Fundamental Physics of the Universe, Via Beirut 2, 34014, Trieste, Italy\\ 
$^{6}$INFN -- Sezione di Trieste\\ 
}

\date{Accepted XXX. Received YYY; in original form ZZZ}

\pubyear{2021}

\begin{document}
\label{firstpage}
\pagerange{\pageref{firstpage}--\pageref{lastpage}}
\maketitle

\begin{abstract}
We present a perturbative treatment of nonlinear galaxy clustering in the context of the cubic Galileon modified gravity model, in terms of 2nd order Lagrangian Perturbation theory and an extension of ellipsoidal collapse that includes Vainshtein screening. We numerically implement such prescriptions in the approximate {\sc pinocchio} code, and use it to generate realisations of the matter density field and halo catalogues with different prescriptions for ellipsoidal collapse. We investigate the impact of three different approximations in the computation of collapse times on the halo mass function, halo bias and matter power spectrum. 
In the halo mass function, both the modified gravity effect and the screening effect are significant in the high mass end, similar to what is found for other MG models. 
We perform a comparison with N-body simulations to assess the validity of our approach, and show that we can reproduce  the same trend observed in simulations for all quantities considered. With a simple modification to the grouping algorithm of {\sc pinocchio} to take into account the gravity modification, and without the need to re-calibrate the algorithm, we show that we can reproduce the linear halo bias and the mildly-nonlinear matter power spectrum of simulations with good accuracy, especially for the implementation with Vainshtein screening. We stress that, while approximate, our method is orders of magnitude faster than a full N-body simulation, making it an optimal tool for the quick generation of large sets of halo catalogues for cosmological observables.
\end{abstract}


\section{Introduction}

Within the standard $\Lambda$CDM cosmological model, the observed accelerated expansion of the Universe is ascribed to the presence of dark energy (DE) in the form of a cosmological constant $\Lambda$: a perfect fluid with constant energy and equation of state, whose negative pressure $P_{\Lambda}=-\rho_{\Lambda}$ is responsible for the accelerated expansion. 
DE makes up most of the energy density of the Universe today together with cold dark matter (CDM), however, its nature remains elusive. In the standard framework, the cosmological constant $\Lambda$ is thought to be related to vacuum energy: an interpretation that, despite the model fitting extremely well most cosmological observations \citep{BOSS:2017, Planck:2018vyg, DES:2021wwk}, poses severe theoretical problems \citep{Carroll:2001}. Additionally, tensions in some of the cosmological parameters as measured by early and late time probes such as the Cosmic Microwave Background and redshift-space distortions of galaxies \citep{Planck:2018vyg,Heymans:2020gsg,DES:2021wwk} have prompted the study of structure formation in the context of beyond-$\Lambda$CDM models. 
Many alternatives have been proposed, from  exotic DE models to modified gravity (MG, see \cite{Bull:2015stt, Koyama:2018, Ishak:2018his} for recent reviews). While the former are mainly characterized by a dynamically evolving equation of state, the latter focus on the possibility that General Relativity (GR) is not the correct theory to describe gravity on cosmological scales, and introduce an additional fifth force that drives cosmic acceleration. 

On the other hand GR has successfully passed several stringent tests, therefore any viable MG model should be able to evade such constraints and reduce to GR on small scales. In order to be consistent with both small- and large-scale observations, the additional MG fifth force has to be shielded in high density regions. This is achieved by means of a so-called screening mechanism. Screening mechanisms can be classified in three general categories \citep{Lombriser:2016zfz}, based on the condition that the Newtonian gravitational potential $\Psi_{\rm N}$ or its derivatives exceed a certain threshold $\Lambda_{\rm T}$:

(i) screening at large field values, such as the chameleon \citep{Khoury:2003aq} or symmetron \citep{Hinterbichler:2010es} models. These are very similar to the large field inflation model: the screening effect turns on in regions where the Newtonian gravitational potential exceeds a certain threshold, $|\Psi_{\rm N}|>\Lambda_{\rm T}$. A high density environment plays a key role in this mechanism;

(ii) screening with first derivatives, such as in k-mouflage models \citep{Babichev:2009ee}. This screening effect operates when the local gravitational acceleration is above a certain threshold, $|\nabla\Psi_{\rm N}|>\Lambda_{\rm T}^2$;

(iii) screening with second derivatives, such as Vainshtein screening \citep{Vainshtein:1972sx}. This screening mechanism activates when the local curvature is large, $|\nabla^2\Psi_{\rm N}|>\Lambda_{\rm T}^3$. Unlike the chameleon mechanism, Vainshtein screening does not rely on the environment, the screening radius is jointly determined by the Schwarzschild radius of the object and the Hubble radius.  

The investigation of alternatives to the standard cosmological model is indeed one of the key targets of modern cosmology. Upcoming Stage-IV galaxy surveys, such as Euclid~\footnote{\url{http://sci.esa.int/euclid}}, LSST~\footnote{\url{http://www.lsst.org}}, WFIRST~\footnote{\url{https://wfirst.gsfc.nasa.gov}}, DESI~\footnote{\url{https://www.desi.lbl.gov}}, J-PAS~\footnote{\url{http://www.j-pas.org/wiki/index.php/Main_Page}}, CSST~\footnote{\url{http://www.bao.ac.cn/csst/}}, will probe the clustering of galaxies to high precision, allowing to test different MG models in the linear and mildly nonlinear regime of structure formation \citep{EuclidTheoryWorkingGroup:2012gxx,Alam:2020jdv}.

It is therefore key that accurate theoretical modelling beyond GR is prepared in advance in order to compare to observations, with a particular focus on the nonlinear and mildly nonlinear regimes. The most reliable tool to trace the growth of structures deep into the nonlinear regime are N-body simulations. However, N-body simulations are computationally extremely expensive, in particular when they include MG. 

This paper studies the modelling of nonlinearities in the framework of perturbation theory (PT) in the context of the cubic Galileon model \citep{Nicolis:2008in}. In particular, we focus on a numerical implementation in the {\sc pinocchio} code (PINpointing Orbit-Crossing Collapsed Hierarchical Objects, \cite{Monaco:2001jg,Monaco:2001jf,Taffoni:2001jh,Monaco:2013qta,Munari:2016aut}), an algorithm to quickly generate simulated dark matter halo catalogues based on Lagrangian PT and ellipsoidal collapse. 

We organize the papers as follows: in section \ref{sec:cubic} we describe the background, linear and nonlinear perturbation evolution of the cubic Galileon model. In section \ref{sec:LPT}, we present the 1st and 2nd order Lagrangian PT for the cubic Galileon model. In section \ref{sec:pinocchio}, we give the prescription of the ``{\sc g3-pinocchio}'' algorithm, including the extension of ellipsoidal collapse. In section \ref{sec:results}, we show the nonlinear matter power spectrum and mass function obtained with the code, as well as a prediction for the linear bias. Finally, we present our conclusions in Section \ref{sec:con}.

\section{Cubic Galileon}
\label{sec:cubic}

Galileon gravity \citep{Nicolis:2008in} is proposed in the inspiration of flat space quantum field theories. 
The latter are invariant under the transformation \(\phi\rightarrow\phi+b_\mu x^{\mu} + c\), where \(b_\mu\) and $c$ are a constant vector and scalar in flat space respectively. The existence of $b_{\mu}$ and $c$ indicates respectively the Galilean and shift symmetry. To avoid the Ostrogradski instability, the field equation includes up to second order time derivatives. When extending to curved space-time, the non-minimal coupling between the Galileon field and the metric breaks the Galilean symmetry, but keeps the shift symmetry preserved. Such model, called covariant Galileon model, has a Lagrangian of the form 
\begin{eqnarray}
\label{eq:lagrangian}
   S&=&\int d^{4}x\sqrt{-g}\left\{ \frac{M_{\rm pl}^{2}}{2}R-\frac{1}{2}c_2X+\frac{c_3}{M^{3}}X\Box\phi+
\frac{c_{4}}{4M^{6}}X^{2}R\right.\nonumber\\ 
&&-\frac{c_{4}}{M^{6}}X[(\Box\phi)^{2}-\phi^{;\mu\nu}\phi_{;\mu\nu}]+
 \frac{3c_{5}}{4M^{9}}X^{2}G_{\mu\nu}\phi^{;\mu\nu}\\
 &&\left.+\frac{c_{5}}{2M^{9}}X[(\Box\phi)^{3}-3\Box\phi\phi_{;\mu\nu}\phi^{;\mu\nu}+2\phi^{;\mu\nu}\phi_{;\mu\sigma}\phi^{;\sigma}_{;\nu}]\right\}\;,\nonumber
\end{eqnarray}
where $M_{\rm pl}$ is the Planck mass \(M_{\rm pl}^{-2}=8\pi G\), \(g, R, G_{\mu\nu}\) are respectively the determinant of the metric, the Ricci scalar and the Einstein tensor, \(c_2, c_3, c_4, c_5\) are dimensionless
constants and \(M^3=M_{\rm pl}H_0^2\), with \(H_0\) the present Hubble parameter. \(X=\phi_{;\mu}\phi^{;\mu}\) is the kinetic term of the scalar field. The semicolon in Eq.~(\ref{eq:lagrangian}) represents the covariant derivative.
There are three branches of covariant Galileon: the G3
branch, known as cubic Galileon, with \(c_3 \neq 0\), \(c_4 = c_{5}=0\), the G4 branch, also called quartic Galileon, with \(c_3,~c_4 \neq 0\), \(c_5=0\) and the G5 branch, also known as quintic Galileon, with \(c_3,~c_4,~c_5 \neq 0\). 
Some recent studies that focus on constraining the covariant Galileon model by means of various cosmological observations include \cite{Barreira:2012kk,Barreira:2013xea,Barreira:2014jha,Neveu:2013mfa,Peirone:2017vcq,Renk:2017rzu,Frusciante:2019puu}. Although this model is disfavored by current CMB and galaxy clustering data, its extension can fit current cosmological observations~\citep{Peirone:2019aua,Frusciante:2019puu}. 

The breakthrough discovery of gravitational waves (GW) from the merger of a neutron star binary \citep{TheLIGOScientific:2017qsa} by the LIGO and Virgo collaborations puts a tight constraint on the GW speed \citep{LIGOScientific:2017zic}, effectively ruling out several MG models. In particular, the G4 and G5 branches are ruled out \citep{Creminelli:2017sry,Ezquiaga:2017ekz,Baker:2017hug} with a high confidence level.
In the G3 branch however, the scalar field is minimally coupled to gravity, and the GW speed is unaltered. Hence, G3 is still a viable MG model even according to GW costraints.

An additional reason to investigate the cubic Galileon model is related to the screening mechanism it features. Since the G3 field drives cosmic acceleration via the non-canonical kinetic energy, the extra gravitational force is screened via the Vainshtein mechanism. While chameleon screening has been extensively discussed during the past decades, in particular in the context of $f(R)$ gravity (for example in \cite{Li:2011vk,Puchwein:2013lza,Llinares:2013jza}), Vainshtein screening is usually studied as feature of nDGP modified gravity (e.g. \cite{Winther:2015wla, Barreira2015, Winther2017, Aguayo2021}).  Studies of this type of screening mechanism that focus on cubic Galileon are instead more rare; some examples include the work of~\cite{Schmidt:2009sg,Barreira:2013eea,Li:2013nua}. 

A first step to include MG in the {\sc pinocchio} code focused indeed on $f(R)$ gravity, which features scale-dependent growth and chameleon screening~\citep{Moretti:2019bob}. 
In this work, we present an implementation of scale-independent MG models with Vainshtein screening, focusing in particular on the G3 model, and study its performance with {\sc pinocchio}. 

\subsection{Background evolution under cubic Galileon}

The Einstein equation and the scalar field equation can be obtained by varying the action with respect to the metric and scalar field respectively. In the spatially flat Friedmann-Lema\^\i tre-Robertson-Walker (FLRW) metric, one can
get the 1st- and 2nd- Friedmann equations
\begin{eqnarray}
3H^{2}&=&\kappa(\bar{\rho}_{\rm m}+\bar{\rho}_{\rm r}+\bar{\rho}_{\phi})\;,\\
0&=&3\dot H+3H^{2}+\frac{\kappa}{2}[\bar{\rho}_{\rm m}+\bar{\rho}_{\rm r}+\bar{\rho}_{\phi}+3(\bar{P}_{\phi}+\bar{P}_{\rm r})]\;,
\end{eqnarray}
where \(\kappa=8\pi G\), \(\bar P_{\rm r}=\frac{1}{3}\bar \rho_{\rm r}\).
Quantities with a bar are background quantities and the over dot represents the time derivative. 
The expressions for \(\bar P_{\phi},\bar{\rho}_\phi \) are
\begin{eqnarray}
    \bar P_{\phi}&=&\frac{1}{2}c_2\dot{\bar\phi}^{2}-2\frac{c_3}{M^{3}}\dot{\bar\phi}^{2}\ddot{\bar\phi}\;,\\
    \bar\rho_{\phi}&=&\frac{1}{2}c_2\dot{\bar\phi}^{2}+6\frac{c_3}{M^{3}}H\dot{\bar\phi}^{3}. 
\end{eqnarray}
The matter component satisfies the continuity equation \(\dot{\bar\rho}_{\rm m}+3H\bar\rho_{\rm m}=0\), the
radiation component satisfies \(\dot{\bar\rho}_{\rm r}+4H\bar\rho_{\rm r}=0\), and the scalar field component (or the DE component) satisfies
\(\dot{\bar\rho}_\phi+3H(\bar\rho_\phi+\bar P_\phi)=0\).
For the scalar field, we use the tracker solution of \cite{DeFelice:2010pv}:
\begin{equation}
\dot{\bar\phi}=\xi H_{0}^{2}/H\;.
\label{eq:tracker}
\end{equation}
The meaning of tracker denotes that different initial conditions of the background Galileon field give rise to different time evolution that eventually merge into a common trajectory.
Here \(\bar \phi\) is the background field, \(H\) is the Hubble parameter and \(\xi\) is a dimensionless constant. Eq.~(\ref{eq:tracker}) expresses the solution of the background scalar field in terms of the Hubble parameter. Naively, it would seem that there are three extra parameters in the G3 model compared with $\Lambda$CDM, namely \(\{c_2,c_3,\xi \}\). 
However, that is not the case: in what follows we show that the number of parameters in G3 is actually the same as those in $\Lambda$CDM. 
By substituting the tracker solution Eq.~(\ref{eq:tracker}) into the 1st Friedmann equation, one can get
\begin{equation}
    E^{4}=(\Omega_{\rm m,0} a^{-3} + \Omega_{\rm r,0} a^{-4}) E^{2} + \frac{1}{6} c_2 \xi^{2} + 2 c_3 \xi^{3}\;, 
\end{equation}
where $E$ is the dimensionless Hubble parameter \(E=H/H_0\), \(\Omega_{\rm m, 0}=\bar\rho_{\rm m,0}/(3M_{\rm pl}^{2}H_0^{2})\) and \(\Omega_{\rm r,0}=\bar\rho_{\rm r,0}/(3M_{\rm pl}^{2}H_0^{2})\) are the present matter density and radiation density respectively and $a$ is the scale factor. Defining
\begin{equation}
\label{eq:omegaL}
  \Omega_{\Lambda,0}=1-\Omega_{\rm m,0}-\Omega_{\rm r,0}=\frac{1}{6}c_2\xi^{2}+2c_3\xi^{3}\;,
\end{equation} 
we can get a parameterised Hubble parameter 
\begin{equation}
    H=H_{0}\sqrt{\frac{\Omega_{\rm m,0}a^{-3}+\Omega_{\rm r,0}a^{-4}+\sqrt{(\Omega_{\rm m,0}a^{-3}+\Omega_{\rm r,0}a^{-4})^{2}+4\Omega_{\Lambda,0}}}{2}}\;.
\end{equation} 
To avoid the scaling degeneracy and without loss of generality~\citep{Barreira:2014jha}, one can assume $c_2=-1$.
Combining the scalar field equation and the tracker solution, we can get a constraining equation on the G3 parameters:
\begin{equation}
\label{eq:constraining}
  c_2\xi+6c_3\xi^2=0\;.  
\end{equation}
From Eq.~(\ref{eq:omegaL}) and (\ref{eq:constraining}), we have
\begin{equation}
\label{eq:c2xi}
    c_3=\frac{1}{6\sqrt{6\Omega_{\Lambda,0}}}, \quad \xi=\sqrt{6\Omega_{\Lambda,0}}\;.
\end{equation}  
One can see that, once $\Omega_{\Lambda,0}$ is given, all the G3 model parameters are fixed and there are no additional parameters with respect to $\Lambda$CDM. Through the 2nd Friedmann equation, we obtain the acceleration parameter
\begin{equation}
    \dot H=\frac{\frac{H_{0}^{4}}{H^{2}}\Omega_{\Lambda,0} - H^{2} - \frac{1}{2} H_{0}^{2} (\Omega_{\rm m,0} a^{-3} + 2 \Omega_{\rm r,0} a^{-4})}
{1+\frac{H_{0}^{4}}{H^{4}}\Omega_{\Lambda,0}}\;. 
\end{equation} 
The above equations fully describe the background cosmology in the G3 model. 
\begin{figure}
	\includegraphics[width=\columnwidth]{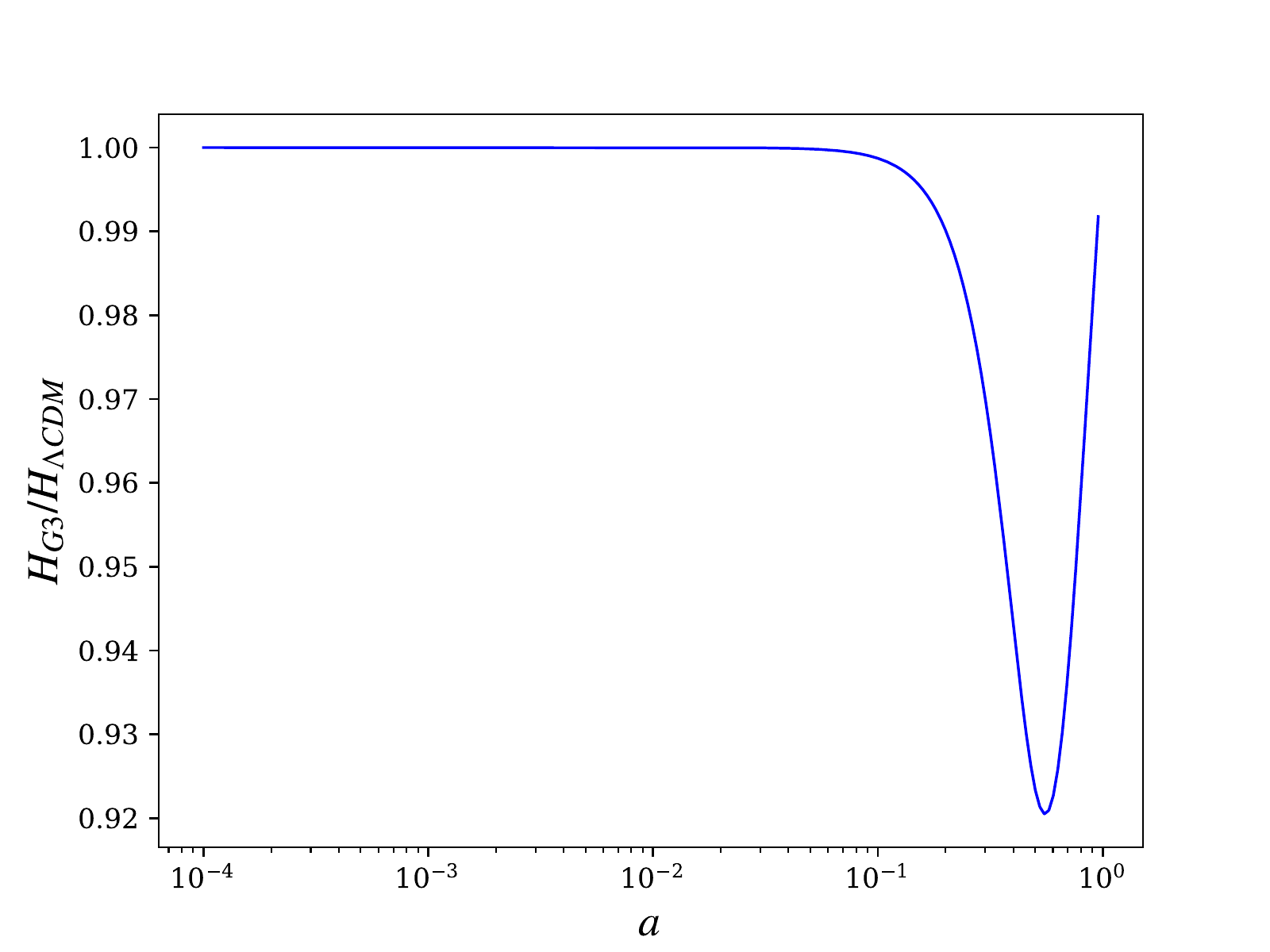}
    \caption{Ratio of the Hubble parameter in G3 and the $\Lambda$CDM model as a function of the scale factor $a$.}
    \label{fig:bkg}
\end{figure}
Figure \ref{fig:bkg} shows the ratio of the Hubble parameter between G3 and \(\Lambda\)CDM as a function of the scale factor $a$: one can see that the two are the same at early times. Starting from $a \simeq 0.1$, the ratio decreases, reaches a minimum around $a \simeq 0.5$, then bounces back and approaches unity at present time.

\subsection{Linear perturbations in cubic Galileon}

In the context of linear perturbation theory and in the framework of GR, the matter anisotropic stress can be ignored. The Weyl potential \(\Phi_+=(\Phi+\Psi)/2\) felt by relativistic particles is equal to the Newtonian potential \(\Psi\) felt by non-relativistic particles, namely \(\Phi_+=\Phi=\Psi\). 
Here we denote the Newtonian potential as \(\Psi\) and the spatial curvature perturbation as \(\Phi\). 
Generally, the equality $\Phi=\Psi$ does not hold in MG models due to the existence of the Compton wavelength of the extra scalar field: the gravitational force below and above this wavelength is different. This phenomenon can be parameterised in the $00$ component of the Einstein equation by means of a function $\mu^{\rm L}$\footnote{We also call this $\mu$ function the {\it gravitational slip function}}, that in general is time- and scale-dependent:
\begin{equation}
    k^2\Psi=-4\pi G\mu^{\rm L}(a,k)a^2\bar\rho_{\rm m}\Delta\;,
\end{equation}
where \(\Delta=\delta+3aHv/k\), with \(\delta=\delta\rho_{\rm m}/\bar\rho_{\rm m}\) the matter density contrast in the Newtonian conformal gauge, and \(v\) the irrotational part of the peculiar velocity.
One can read the $\mu^{\rm L}$ function as the ratio between the effective gravitational coupling $G_{\rm eff}$ and the Newton constant $G_{\rm N}$, with the superscript ``L'' denoting linear level. In general, chameleon models feature a $k$-dependent $\mu^{\rm L}$ function, which translates in scale dependent growth even at linear level. 
For the k-essence type of MG models however, cosmic acceleration is driven by the non-canonical kinetic energy, the scalar field is effectively massless and the corresponding Compton wavelength is on or above the Hubble horizon scale. 
The effective gravitational coupling at linear level may not be equal to the Newton constant, but it is constant on all the scales relevant to the linear regime \citep{Peirone:2017vcq}.  
Hence, for this type of MG models (that include the G3 model we are considering in this work), $\Phi=\Psi$ is still valid and $\mu^{\rm L}$ is a function of time only. 

The linear perturbation regime of MG models has been extensively studied in the past few years. In particular, it has been shown both theoretically and numerically that single field models can be re-expressed in the language of the Effective Field Theory of Dark Energy (EFTofDE, \cite{Gubitosi:2012hu,Bloomfield:2012ff,Piazza:2013coa,2014PhRvD..89j3530H,Zumalacarregui:2016pph,Frusciante:2019xia}).
The action in the EFTofDE reads
\begin{eqnarray}
S&=&\int d^4x\sqrt{-g}\left \{ \frac{M_{\rm pl}^2}{2}\Omega(t)R+\Lambda(t)-c(t)\delta g^{00}\right.\nonumber\\
&&+\frac{M_2^4(t)}{2}(\delta g^{00})^2 -\frac{\bar M_1^3(t)}{2}\delta g^{00}\delta K_{\mu}^\mu-\frac{\bar M_2^2(t)}{2}(\delta K_\mu^\mu)^2 \nonumber \\
&& -\frac{\bar M_3^2(t)}{2}\delta K^i_j\delta K_i^j+
 \frac{\hat M^2(t)}{2}\delta g^{00}\delta R^{(3)}\\ 
&&\left.+m_2^2(t)(g^{\mu\nu}+n^\mu n^\nu)\partial_\mu(g^{00})\partial_\nu(g^{00})\right\} +S_{\rm m}[g_{\mu\nu},\chi_i]\;,\nonumber
\end{eqnarray}
where $\delta g_{00}$, $\delta K_{\mu\nu}$, $\delta K$,
$\delta R^{(3)}$ are perturbations of the time-time component of the metric, the external curvature and its trace, and the 3-dimensional Ricci scalar in the constant time hypersurface, \(S_{\rm m}\) is the minimally coupled term for all matter fields \(\chi_i\) with metric \(g_{\mu\nu}\), and $M_{\rm i}$, $\bar{M_{\rm i}}$, and $\hat{M_{\rm i}}$ are the EFT functions. The perturbation of the scalar field can be expressed by the infinitesimal time diffeomorphism, \(t\rightarrow t+\pi(x^\mu)\), where \(\pi\) is the perturbation of the scalar field.

Following \cite{Pogosian:2016pwr}, we combine the Einstein equation and  the scalar field equation under the quasi-static approximation, which applies only to scales below the sound horizon of the scalar field. Under this approximation, we ignore the time derivatives of the gravitational potential and the scalar field. One can then derive the following expression for \(\mu^{\rm L}(a,k)\) in the EFTofDE framework:
\begin{equation}
 \frac{\mu^{\rm L}}{2M_{\rm pl}^2}=\frac{f_1+f_2a^2/k^2}{f_3+f_4a^2/k^2}\;,
\end{equation}
with
\begin{eqnarray}
    f_1&=&c-\frac{1}{2}(H+\partial_t)\bar M_1^3\;,\nonumber \\
    f_2&=&-3c\dot H+\frac{3}{2}(3H\dot H+\dot H\partial_t+\ddot H)\bar M_1^3\;,\nonumber \\
    f_3&=&2M_{\rm pl}^2\Omega f_1-\frac{1}{2}\bar M_1^6\;,\nonumber \\
    f_4&=&2M_{\rm pl}^2\Omega f_2\;.
\end{eqnarray}

\begin{figure}
	\includegraphics[width=\columnwidth]{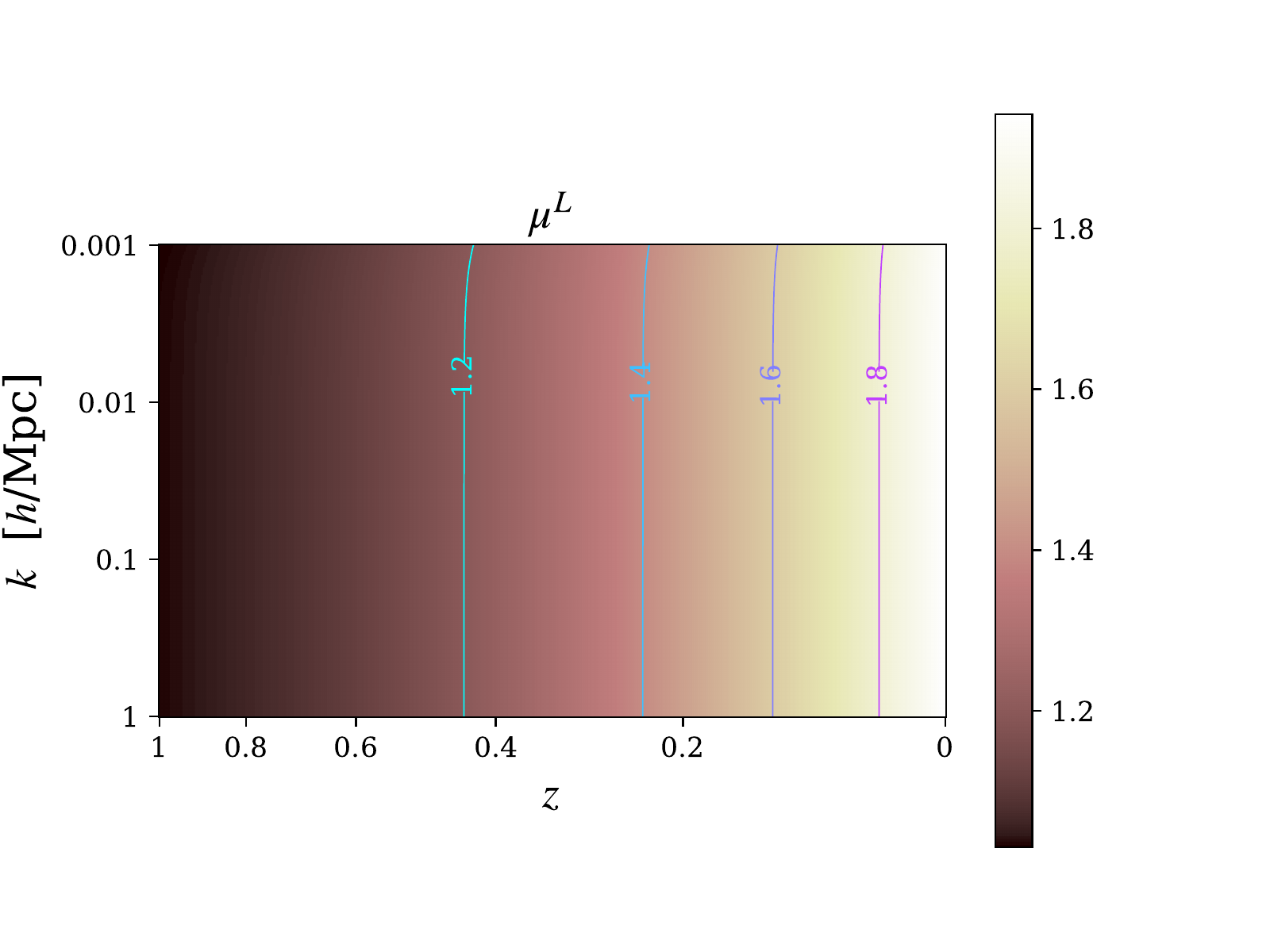} 
	\includegraphics[width=\columnwidth]{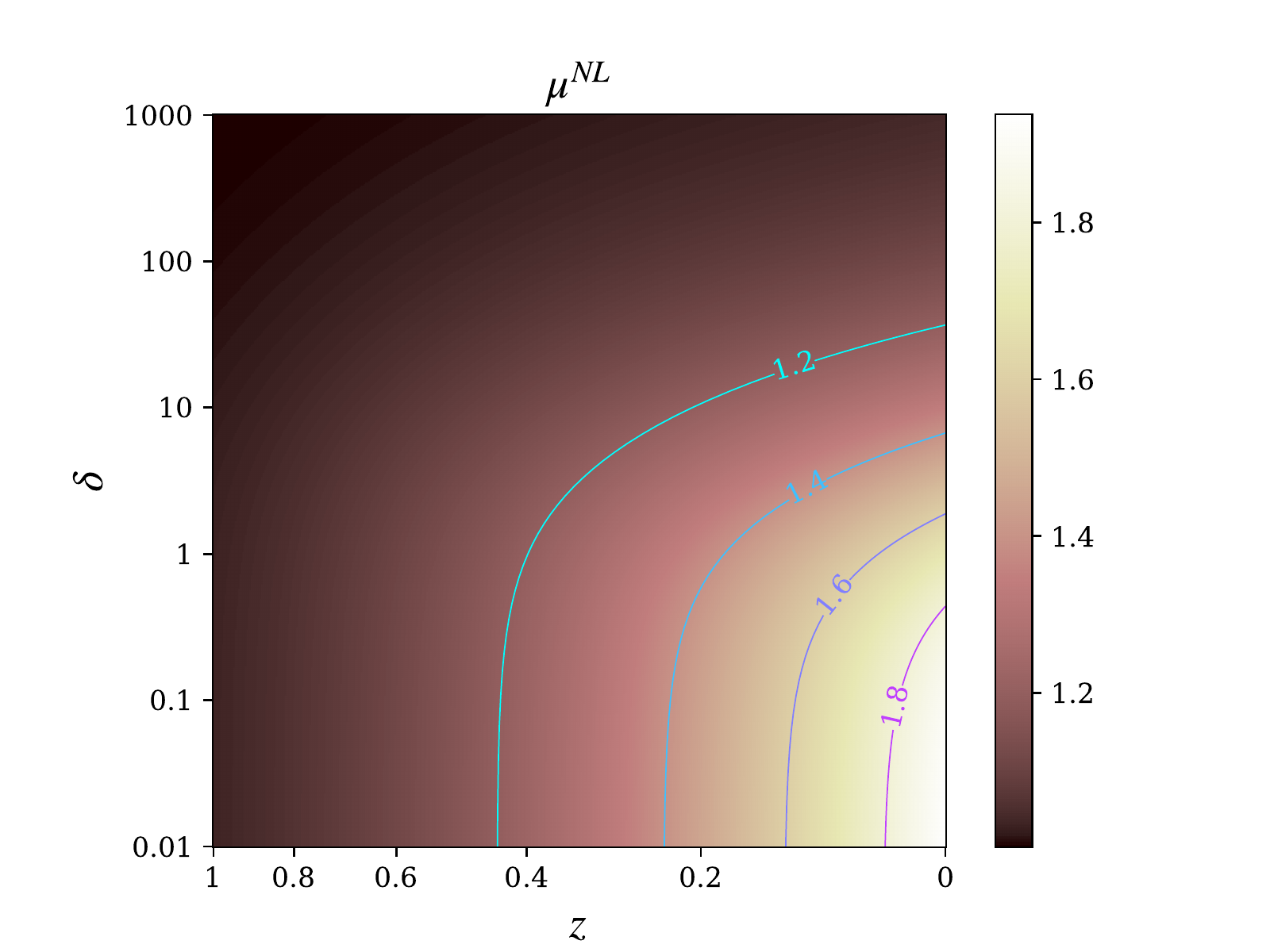}
    \caption{Two dimensional contour plot of the $\mu$ function. The upper panel shows the linear part $\mu^{\rm L}$ as a function of redshift $z$ (horizontal axis) and wavenumber $k$ (vertical axis), with lines showing the isocontours. The lower panel shows the same quantity but including Vainshtein screening $\mu^{\rm NL}$ (Eq.~\ref{eq:munl}), again as a function of redshift $z$ (horizontal axis), but with the local matter density contrast $\delta$ on the vertical axis. Values different from 1 mark a deviation from GR.}
    \label{fig:mu}
\end{figure}

In the upper panel of Figure \ref{fig:mu}, we show the linear part of the $\mu$ function as a function of redshift $z$ and wavenumber $k$. The values are encoded in the color bar. On top of that, we show contours with constant $\mu^{\rm L}$ values. 
One can see that the contours are vertically distributed, in line with the fact that the modification of the gravitational force is scale independent. 
Furthermore, the $\mu$ values are significantly larger at low redshifts: the gravity enhancement reaches a value of around $2$ at present time, while it approaches unity for redshifts $z>1$. 
In the lower panel of Figure \ref{fig:mu} we show the nonlinear $\mu^{\rm NL}$ function that includes Vainshtein screening, discussed in the next section, as a function of both redshift $z$ (horizontal axis) and the local density contrast $\delta$ (vertical axis). One can see that with higher local density, the extra gravitational force is shielded for all redshifts considered.

\subsection{Nonlinear clustering in cubic Galileon}

We now consider the 2nd order perturbation contribution. 
Under the quasi-static approximation, after ignoring the terms suppressed by the time derivatives and Hubble expansion, the 00 component of the Einstein equation reads \citep{Frusciante:2020zfs}
\begin{equation}
\label{eq:00}
    \frac{\partial^{2}\Phi}{a^{2}}=4\pi G\bar\rho_{\rm m}\delta_{\rm m}-8\pi G \frac{c_3}{M^3}\dot{\bar\phi}^{2}\frac{\partial^{2}(\delta\phi)}{a^{2}}\;,
\end{equation}
where \(\partial\) denotes derivatives with respect to comoving spacial coordinates. The equation for the scalar field is 
\begin{equation}
\begin{aligned}
\label{eq:phi} -\frac{c_3}{M^3}\dot{\bar\phi}^2\frac{\partial^{2}\Psi}{a^{2}}=&\left[\frac{1}{2}c_2+2\frac{c_3}{M^3}(\ddot{\bar\phi}+2H\dot{\bar\phi}) \right]\frac{\partial^2\delta\phi}{a^2}+\\
&\frac{c_3}{M^3}\left[(\frac{\partial^2\delta\phi}{a^2})^2-(\frac{\partial_i\partial_j\delta\phi}{a^2})^2\right]\;.
\end{aligned}
\end{equation}
Since \(\Psi=\Phi\) is valid in both the linear and nonlinear regime, combining Eqs.~(\ref{eq:00}) and (\ref{eq:phi}) one gets 
\begin{equation}
\label{eq:phi2}
    \frac{\partial^{2}(\delta\phi)}{a^{2}}+\lambda^{2}(a)\left[(\frac{\partial^{2}(\delta\phi)}{a^{2}})^{2}-(\frac{\partial_{i}\partial_{j}(\delta\phi)}{a^{2}})^{2}\right]=-4\pi G \zeta(a)\bar\rho_{\rm m}\delta_{\rm m}, 
\end{equation}
where
\begin{equation}
\lambda^2(a)=\frac{c_3/M^3}{\frac{1}{2} c_2 + 2 \frac{c_3}{M^3}(\ddot{\bar{\phi}} + 2H\dot{\bar{\phi}}) - 8\pi G(\frac{c_3}{M^3})^2\dot{\bar{\phi}}^4} \; , \; \zeta(a)=\lambda^2\dot{\bar{\phi}}^2\;. \nonumber
\end{equation}
Assuming a spherically symmetric profile for \(\delta\phi\), Eq.~(\ref{eq:phi2}) becomes
\begin{equation}
\frac{1}{r^2}\frac{d}{dr}\left(r^2\frac{d\delta\phi}{dr}\right)-\frac{2\lambda^2}{r^2}\frac{d}{dr}\left[r\left(\frac{d\delta\phi}{dr}\right)^2\right]=-4\pi G\zeta\bar\rho_{\rm m}\delta_{\rm m}\;,
\end{equation}
a second order ordinary differential equation. After integrating once, it becomes 
\begin{equation}
\label{eq:profile}
r^2\frac{d\delta\phi}{dr}-2\lambda^2r\left(\frac{d\delta\phi}{dr}\right)^2=-G\zeta m(r)\;,
\end{equation}
where \(m(r)=4\pi\int_0^rr^{\prime2}\bar\rho_{\rm m}(r^\prime)\delta_{\rm m}(r^\prime)dr^\prime\)
is the mass enclosed in a sphere of radius \(r\). 
Considering a top-hat density profile with radius \(R\), when \(r\leq R\) the physical solution of Eq.~(\ref{eq:profile}) reads \citep{Frusciante:2019puu}
\begin{equation}
\label{eq:dphidr}
\frac{d\delta\phi}{dr}=\frac{r}{4\lambda^2}\Big(1-\sqrt{1+\frac{r_V^3}{r^3}}\Big)\;,
\end{equation} 
where \(r_V\) is the Vainshtein radius of the enclosed mass, given by
\begin{equation}
r_V^3=8Gm(r)\lambda^2\zeta\;. 
\end{equation}
From Eq.~(\ref{eq:dphidr}) one can see that \( d\delta\phi/dr \propto r \) within \( r \leq R \), so
\( \partial^2(\delta\phi)\) is constant. Taking the derivative of Eq.~(\ref{eq:dphidr}) and substituting it in Eq.~(\ref{eq:00}), we get the modified Poisson equation which is constant within the Top Hat
\begin{equation}
    \frac{\partial^2\Psi}{a^2}=4\pi G\mu^{\rm NL}(a,R)\bar\rho_{\rm m}\delta_{\rm m}\;,
\end{equation}
where \(\mu^{\rm NL}(a,R)\) includes the nonlinear contribution
\begin{equation}
\label{eq:munl}
    \mu^{\rm NL}(a,R)=1+2(\mu^{\rm L}-1)\Big(\frac{R}{R_V}\Big)^3\Big(\sqrt{1+\frac{R_V^3}{R^3}}-1\Big).
\end{equation}
In the above equation, \(\mu^{\rm L}=1+8\pi G\frac{c_3}{M^3}\dot{\bar\phi}^2\zeta(a)\) is the linear part of the \(\mu\) function. \(R_V\) is the Vainshtein radius of the enclosed mass with a top-hat density profile \(R_V^3=8G m(R)\lambda^2\zeta \propto R^3\). Then, one can write
\begin{equation}
\label{eq:vainradius}
    \Big(\frac{R}{R_V}\Big)^3=\frac{1}{4\Omega_{\rm m}H^2\lambda^2\zeta}\frac{1}{\delta_{\rm m}}.
\end{equation}
When the matter perturbations are large such that \(R\ll R_V\), \(\mu^{\rm NL}\rightarrow 1\) recovering GR, while for small perturbations when \(R\gg R_V\), \(\mu^{\rm NL}\rightarrow \mu^{\rm L}\), recovering the linear result.
In the lower panel of Figure \ref{fig:mu}, we plot \(\mu^{\rm NL}\) as a function of redshift $z$ and the top-hat object density $\delta$. One can see that for low redshifts \(z<1\), it deviates from GR, same as the linear \(\mu^{\rm L}\) function. 
When the density contrast is about 0.01$\sim$0.1, \(\mu^{\rm NL}\) is similar to \(\mu^{\rm L}\), roughly depending on time only. For large values of \(\delta>1000\), \(\mu^{\rm NL}\rightarrow 1\), recovering GR. At fixed redshift, the value of \(\mu^{\rm NL}\) for a high density contrast is smaller than that of a low density contrast.

\section{Lagrangian perturbation theory for Cubic Galileon}
\label{sec:LPT}

In this section we briefly review the linear and 2nd order Lagrangian perturbation theory (LPT) for the G3 model \cite{2021SSPMA..51g9511S}. The equation of motion for the displacement field ($\vec{S}$) in Lagrangian coordinates are
\begin{equation}
\label{eq:displ}
    \nabla_x\cdot\hat T\vec S=-A(a)\delta-B(a)[(\frac{\partial^{2}(\delta\phi)}{a^{2}})^{2}-(\frac{\partial_{i}\partial_{j}(\delta\phi)}{a^{2}})^{2}]\;,
\end{equation}
where
\begin{eqnarray}
    \hat T &=& \frac{d^{2}}{dt^{2}}+2H\frac{d}{dt}\;,\\
    A(a) &=& 4\pi G\bar\rho_{\rm m}\left(1+8\pi G\frac{c_3}{M^3}\dot{\bar\phi}^{2}\zeta(a)\right)\;, \\
    B(a) &=& 8\pi G\frac{c_3}{M^3}\dot{\bar\phi}^{2}\lambda^2(a). 
\end{eqnarray}
Notice that the differential operation $\nabla_x$ is in Eulerian coordinates. Transforming to Lagrangian coordinates one can write
\begin{equation}
\label{eq:E2L}
    \nabla_x\cdot\hat T\vec S=(J^{-1})_{ij}\hat TS_{i,j}=(\delta_{ij}-S_{i,j})\hat TS_{i,j}=\hat TS_{i,i}-S_{i,j}\hat TS_{i,j}\;,
\end{equation}
where $J$ is the Jacobian of the coordinate transformation, and \(S_{i,i},S_{i,j}\) are now spatial derivatives with respect to Lagrangian coordinates. Substituting Eq.~(\ref{eq:E2L}) in Eq.~(\ref{eq:displ}), we can write the displacement field equation in Lagrangian coordinates
\begin{equation}
    \hat TS_{i,i}-S_{i,j}\hat TS_{i,j}=-A(a)\delta-B(a)[(\frac{\partial^{2}(\delta\phi)}{a^{2}})^{2}-(\frac{\partial_{i}\partial_{j}(\delta\phi)}{a^{2}})^{2}].
\end{equation}
Expanding \(\delta,\delta\phi,\vec S\) with respect to a small parameter $\varepsilon$
\begin{eqnarray}
    \delta &=& \varepsilon \delta^{(1)}+\varepsilon^{2}\delta^{(2)}+..., \\
    \vec{S} &=& \varepsilon\vec{S}^{(1)}+\varepsilon^{2} \vec{S}^{(2)}+...\\
    \delta\phi &=& \varepsilon (\delta\phi)^{(1)}+\varepsilon^{2}(\delta\phi)^{(2)}+..., 
\end{eqnarray}
one can split the above equations into serial differential equations according to their orders. At 1st order, the displacement field equation reads
\begin{equation}
(\hat T-A(a))S_{i,i}^{(1)}(t,\vec q)=0.  
\end{equation}
Since \((\hat T-A(a))\) only depends on time,
\(S_{i,i}^{(1)}(t,\vec q)\) can be separated into time- and space-dependent components. 
Transforming the above equation to Fourier space, we obtain 
\begin{equation}
\label{eq:S1}
    S_{i,i}^{(1)}(\vec k,t)=-D_1(t)\delta^{(1)}(\vec k,t_0)\;,
\end{equation}
where the 1st order growth factor \(D_1(t)\) satisfies 
\begin{equation}
    (\hat T-A(a))D_1(t)=0.
\end{equation}
Like in standard GR, the linear growth in G3 only has temporal dependence. For 2nd order in PT, the displacement field equation reads
\begin{eqnarray}
\label{eq:displ2}
    &&\hat TS_{i,i}^{(2)}(\vec k,t)-[S_{i,j}^{(1)}\hat TS_{i,j}^{(1)}](\vec k,t)=
    \\ 
    && -A(a)\delta^{(2)}(\vec k,t)-B(a)\left[(\frac{\partial^{2}(\delta\phi)^{(1)}}{a^{2}})^{2}-(\frac{\partial_{i}\partial_{j}(\delta\phi)^{(1)}}{a^{2}})^{2}\right](\vec k,t). \nonumber
\end{eqnarray}
Through Eq.~(\ref{eq:phi2}) we obtain
\begin{equation}
\label{eq:dphi1}
    \frac{k^2}{a^2}(\delta\phi)^{(1)}(\vec k,t) = 4\pi G\zeta(a)\bar\rho_{\rm m}D_1(t)\delta^{(1)}(\vec k,t_0). 
\end{equation}
Plugging Eq.~(\ref{eq:dphi1}), (\ref{eq:S1}) into Eq.~(\ref{eq:displ2}), we can write
\begin{eqnarray}
\label{eq:S2}
    &&[\hat T-A(a)]S_{i,i}^{(2)}(\vec k,t) = -\left[\frac{1}{2}A(a)+B(a)(4\pi G\bar\rho_{\rm m}\zeta(a))^2\right]\times\nonumber \\
    &&\int_{\vec k_{12}=\vec k}\left[1-\frac{(\vec k_1\cdot\vec k_2)^2}{k_1^2k_2^2}\right]D_1^2(t)\delta_{k_1}^{(1)}(t_0)\delta_{k_2}^{(1)}(t_0)\\
    &&=-C(a)\int_{\vec k_{12}=\vec k}\left[1-\frac{(\vec k_1\cdot\vec k_2)^2}{k_1^2k_2^2}\right]D_1^2(t)\delta_{k_1}^{(1)}(t_0)\delta_{k_2}^{(1)}(t_0)\nonumber\;,
\end{eqnarray}
where \(\int_{\vec k_{12}=\vec k}\) is short-hand notation for
\(\int\frac{d^3\vec k_1d^3\vec k_2}{(2\pi)^3}\delta(\vec k-\vec k_1-\vec k_2)\), and \[
C(a)=\frac{1}{2}A(a)+B(a)(4\pi G\bar\rho_{\rm m}\zeta(a))^2.\] 
The 2nd order displacement field then reads 
\begin{equation}
\label{eq:Sii}
S_{i,i}^{(2)}(\vec k,t)= 
\int_{\vec k_{12}=\vec k}\widetilde D_2(\vec k_1,\vec k_2,t)D_1^2(t)\delta_{k_1}^{(1)}(t_0)\delta_{k_2}^{(1)}(t_0), 
\end{equation}
where \(\widetilde D_2(\vec k_1,\vec k_2,t)\) is a normalized second order growth factor that satisfies 
\begin{equation}
\label{eq:D21}
[\hat T-A(a)]\widetilde D_2(\vec k_1,\vec k_2,t)=-C(a)\left[1-\frac{(\vec k_1\cdot\vec k_2)^2}{k_1^2k_2^2}\right]. 
\end{equation}
Since \([\hat T-A(a)]\) depends only on time, \(\widetilde D_2(\vec k_1,\vec k_2,t)\) can also be separated into the product of time component \(\widetilde D_2(t)\) and
space component \([1-\frac{(\vec k_1\cdot\vec k_2)^2}{k_1^2k_2^2}]\).
Thus Eq.~(\ref{eq:Sii}) can be rewritten as 
\begin{equation}
\label{eq:d2convl}
S_{i,i}^{(2)}(\vec k,t)= 
D_1^2(t)\widetilde D_2(t)\int_{\vec k_{12}=\vec k}\left[1-\frac{(\vec k_1\cdot\vec k_2)^2}{k_1^2k_2^2}\right]\delta_{k_1}^{(1)}(t_0)\delta_{k_2}^{(1)}(t_0)\;. 
\end{equation}
Redefining \(-2D_1^2(t)\widetilde D_2(t)\) as \(D_2(t)\), then the second order growth factor
\(D_2(t)\) satisfies
\begin{equation}
[\hat T-A(a)]D_2(t)=2C(a)D_1^2(t)\;. 
\end{equation}

Again, \(D_2\) is independent of \(k\), as it is in the standard GR case. To get the second order displacement field \(S_{i,i}^{(2)}(\vec k,t)\), one just
needs to convolve the initial linear density fields with a kernel as in Eq.~(\ref{eq:d2convl}). 
The difference between G3 and $\Lambda$CDM lays in the temporal evolution of the linear and 2nd order growth factors.
When \(A(a)=4\pi G\bar\rho_{\rm m},B(a)=0,C(a)=2\pi G\bar\rho_{\rm m}\), the equations for the 1st- and 2nd- order displacements recover \(\Lambda\)CDM~\citep{2021SSPMA..51g9511S}.

\begin{figure}
	\includegraphics[width=\columnwidth]{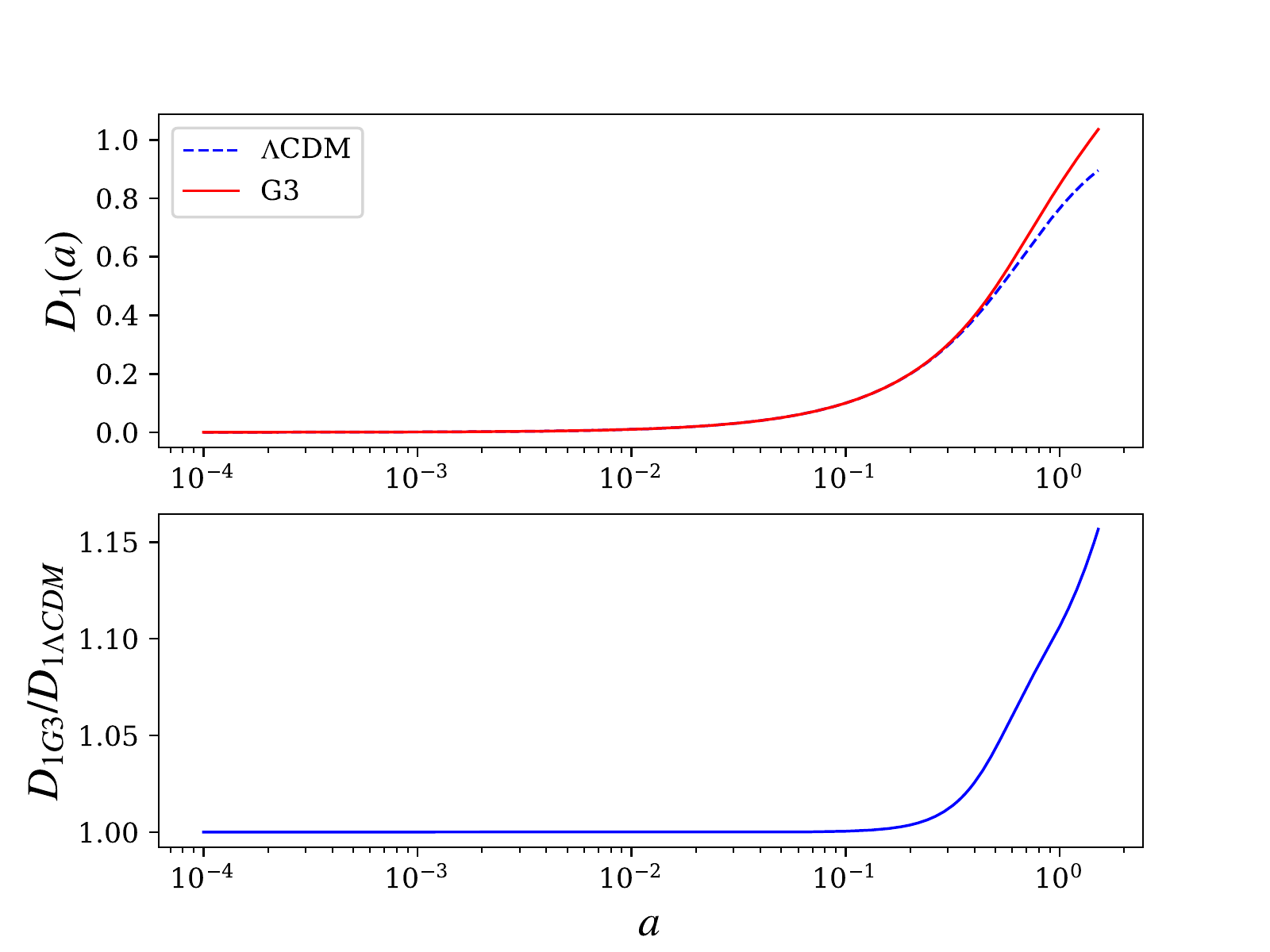} 
	\includegraphics[width=\columnwidth]{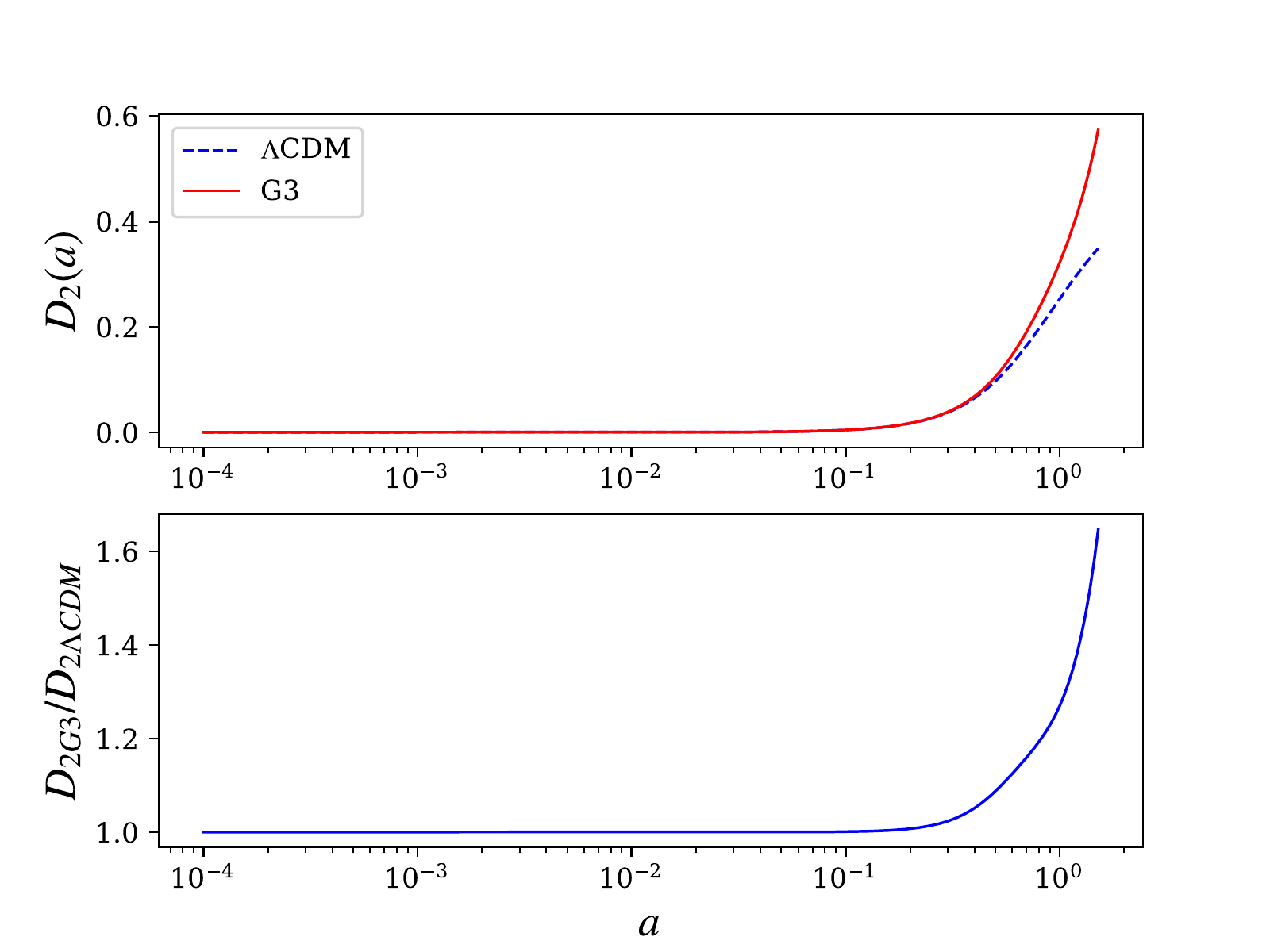}
    \caption{The linear (top panel) and 2nd order (third panel) growth factors as a function of the scale factor $a$. Dashed blue lines are $\Lambda$CDM quantities, while solid red lines are G3 quantities. The second and bottom panels show respectively the ratio of the G3 linear and 2nd order growth functions to their $\Lambda$CDM counterparts.}
    \label{fig:growth}
\end{figure}

\begin{figure}
	\includegraphics[width=\columnwidth]{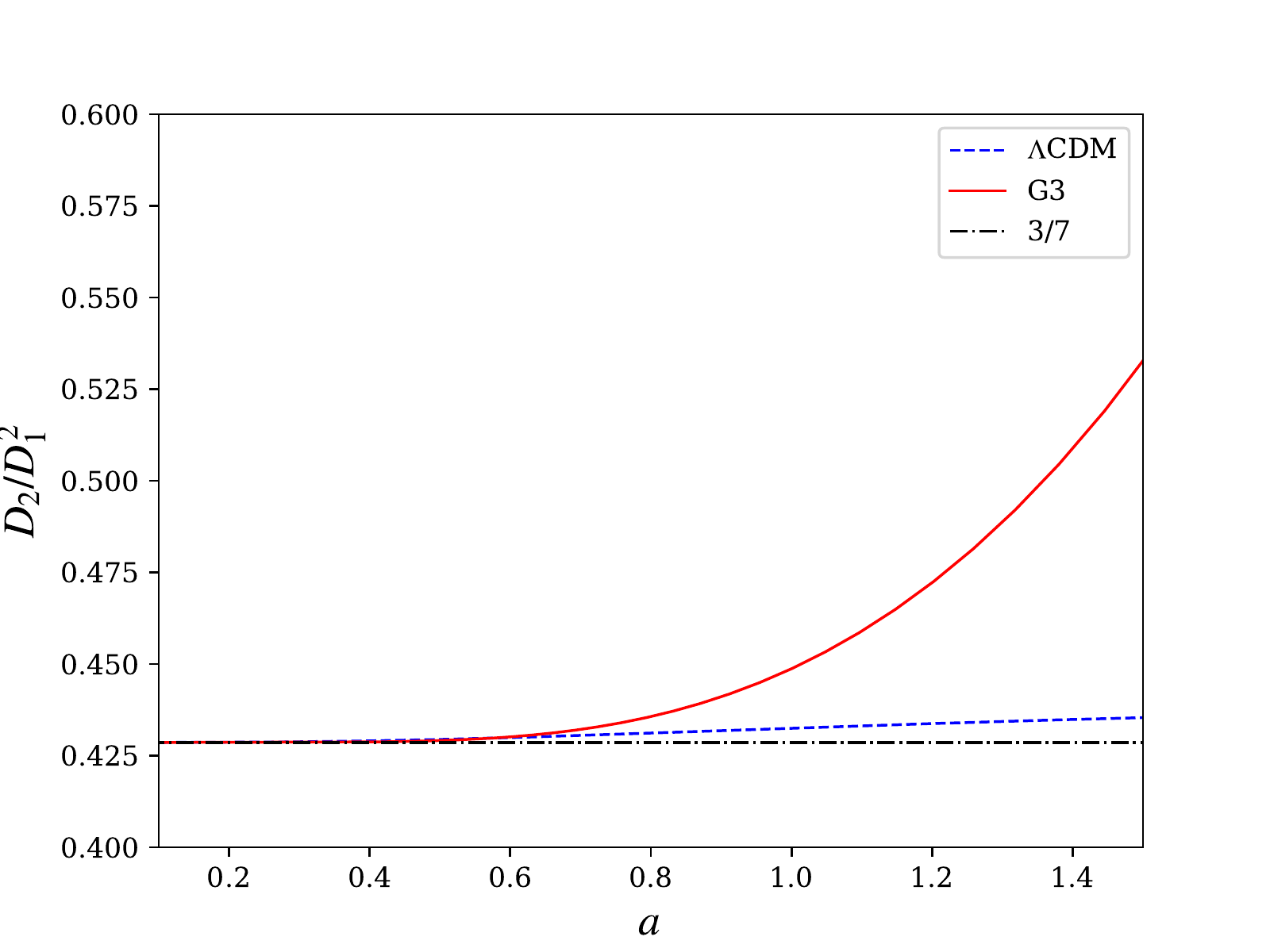}
    \caption{The ratio $D_2/D_1^2$ as a function of the scale factor $a$. The black dot-dashed line shows the Einstein-de Sitter value of $3/7$; while the $\Lambda$CDM (blue dashed) line is close to it, the G3 (solid red) deviates substantially.}
    \label{fig:growth2}
\end{figure}

We choose the initial conditions for the $D_1$ differential equation to match an Einstein-de Sitter universe, i.e. \(D_1(a)=a,\frac{dD_1}{da}=1\). 
In the top panel of Fig.~\ref{fig:growth} we show \(D_1(a)\) in G3 (red solid line) and $\Lambda$CDM (blue dashed line), with their ratio in the second panel, as a function of the scale factor $a$.
One can see that \(D_1\) in G3 is the same as in \(\Lambda\)CDM at early times. Starting from $a\simeq 0.2$, \(D_1\) in G3 increases faster than in \(\Lambda\)CDM, with their ratio reaching $\sim 15\%$ at present times. 
The initial conditions for \(D_2\) are assumed to be \(D_2(a)=\frac{3}{7}a^2,\frac{dD_2}{da}=\frac{6}{7}a\), corresponding to matter domination.
In the third panel of Figure \ref{fig:growth} we show \(D_2(a)\) in G3 (red solid line) and $\Lambda$CDM (blue dashed line), with their ratio in the bottom panel, as a function again of the scale factor $a$. 
Similarly to \(D_1\), at early times \(D_2\) for G3 and \(\Lambda\)CDM are the same. Moreover, at $a\simeq 0.2$, \(D_2\) in G3 increases much faster than in $\Lambda$CDM. 
The difference is more sizable than in $D_1$, about $60\%$ for $a=1$.
Since the growth functions are enhanced, cubic Galileon predicts stronger galaxy clustering than \(\Lambda\)CDM. In Figure \ref{fig:growth2}, we show the ratio of \(D_2/D_1^2\). This ratio is exactly $3/7$ in the Einstein-de Sitter universe.
The extra clustering power in G3 comes from a larger effective gravitational constant $G_{\rm eff}$ compared to the standard Newton constant $G_{\rm N}$.

\section{Implementation of G3 in {\sc pinocchio}}
\label{sec:pinocchio}

The {\sc pinocchio} code (PINpointing Orbit-Crossing Collapsed Hierarchical
Objects) \cite{Monaco:1997cq,Monaco:2001jg,Monaco:2001jf,Taffoni:2001jh,Monaco:2013qta,Chuang:2014toa,Munari:2016aut,Rizzo:2016mdr} is a semi-analytical algorithm for generating realisations of hierarchical formation history of dark matter halos. It is based on LPT and ellipsoidal collapse.
The code evolves the initial linear density field of a
given primordial power spectrum on a regular grid in Fourier space. Then, particles are displaced according to Lagrangian PT. For cubic Galileon, the 1st and 2nd order initial displacement fields in Fourier space are given by
\begin{eqnarray}
\label{eq:displs}
    &&S_{i,i}^{(1)}(\vec k,a_0)=-\delta^{(1)}(\vec k,a_0)\;,\\ 
    &&S_{i,i}^{(2)}(\vec k,a_0)=-\frac{1}{2} \int_{\vec k_{12}=\vec k}\left[1-\frac{(\vec k_1\cdot\vec k_2)^2}{k_1^2k_2^2}\right]\delta_{k_1}^{(1)}(a_0)\delta_{k_2}^{(1)}(a_0)\;.\nonumber\\
\end{eqnarray}
The initial time \(a_0\) is arbitrary and related to the normalisation choice for the linear power spectrum, {\sc pinocchio} chooses to normalise at \(a_0=1\) so that \(D_1(a_0)=1\). For convenience, one can define normalised growth functions
\begin{equation}
D_1^\prime(a)=\frac{D_1(a)}{D_1(a=1)},\quad D_2^\prime(a)=\frac{D_2(a)}{D_1^2(a=1)}\;.
\end{equation}
Then the positions of dark matter particles then are given by
\begin{equation}
\label{eq:particle}
\vec x(a)=\vec q+D_1^\prime(a)\vec S^{(1)}(\vec q,a_0)+D_2^\prime(a)\vec S^{(2)}(\vec q,a_0)\;, \end{equation} 
where $\vec S^{(i)}(\vec q,a)$ are the Fourier transforms of the displacements of Eq.~\ref{eq:displs}.

Ellipsoidal collapse is used to predict the collapsing time of each dark matter particle. The dynamics of ellipsoidal collapse can be completely described by following the evolution of the three principal axes of the ellipsoid \citep{1996ApJS..103....1B}. The physical length of ellipsoid semi axes is \(r_i=a_iq\),
where $q$ is the comoving radius of the ``Lagrangian sphere'' (a sphere concentric with and sharing the same mass as the ellipsoid, but with density equal to  the background density) and \(a_i\) denotes the time evolution of each axis. By enforcing mass conservation we can write \(a^3q^3\bar\rho_{\rm m}=a_1a_2a_3q^3\rho_{\rm m,e}\), where $\rho_{\rm m,e}$ is the density within the ellipsoid, so that the nonlinear overdensity $\delta$ is
\begin{equation}
\delta=\frac{a^3}{a_1a_2a_3}-1\;.
\end{equation} 
Following the approach of \cite{1996ApJS..103....1B}, we can derive the evolution for the \(a_i\) for the G3 model: 
\begin{equation}
\label{eq:ellip}
\frac{d^2a_i}{dt^2}=\frac{\ddot a}{a}a_i-4\pi G\mu\bar\rho_{\rm m}\left[\frac{\delta}{3}+\frac{5}{4}b_i^\prime(t)+\frac{b_i^\prime(t)}{2}\delta\right]a_i\;,
\end{equation}
where
\begin{equation}
\label{eq:int}
    b_i^\prime(t)=-\frac{2}{3} + a_1(t)a_2(t)a_3(t)\int^{\infty}_{0}\frac{d\tau}{[a_{i}^{2}(t)+\tau]\Pi^{3}_{j=1}(a_{j}^{2}(t)+\tau)^{1/2}} 
\end{equation}
and the deviation from $\Lambda$CDM is encoded in the $\mu$ function discussed in the previous sections.
Eq.~(\ref{eq:ellip}) is an integro-differential equation, which should in principle be solved for each particle, resulting in a computationally expensive prescription. In its standard version, {\sc pinocchio} relies on an approximation based on LPT to compute collapse times. Such an approximation however does not hold in the case of models where growth is substantially different than $\Lambda$CDM. For our purposes, we adopt the description of the ellipsoid's evolution presented in \cite{Nadkarni-Ghosh:2014bwa}.
This is equivalent to the \cite{1996ApJS..103....1B} approach, but avoids the integral of Eq.~\ref{eq:int} by resorting to a set of nine dimensionless parameters:
\begin{eqnarray}
    \lambda_{\rm a,i}&=&1-\frac{a_{\rm i}}{a}\;, \\
    \lambda_{\rm v,i}&=&\frac{1}{H}\frac{\dot a_{\rm i}}{a_{\rm i}}-1\;,\\
    \lambda_{\rm d,i}&=&\frac{\delta}{3}+\frac{5}{4}b_i^\prime(t)+\frac{b_i^\prime(t)}{2}\delta.
\end{eqnarray}
Here \(\lambda_{\rm a,i}\) correspond to the eigenvalues of the deformation tensor and characterize the shape of the ellipsoid, \(\lambda_{\rm v,i}\) describe the deviation of the velocity of the \(i\)th axis from the background Hubble flow and \(\lambda_{\rm d,i}\) correspond to the eigenvalues of the tensor of
second derivatives of the gravitational potential. Taking the time
derivative of \(\lambda_{\rm a,i},\lambda_{\rm v,i},\lambda_{\rm d,i}\), we obtain the second order ordinary differential equations
\begin{eqnarray}
\label{eq:NGS}
    &&\frac{d\lambda_{\rm a,i}}{d\ln a}=-\lambda_{\rm v,i}(1-\lambda_{\rm a,i})\;,\\
    &&\frac{d\ln\lambda_{\rm v,i}}{d\ln a}=-\frac{3}{2}\mu\Omega_{\rm m}\lambda_{\rm d,i}-\lambda_{\rm v,i}(2+\frac{\dot H}{H^2})-\lambda_{\rm v,i}^2\;,\nonumber\\
    &&\frac{d\lambda_{\rm d,i}}{d\ln a}=-(1+\delta)(\delta+\frac{5}{2})^{-1}(\lambda_{\rm d,i}+\frac{5}{6})\sum_{j=1}^{3}\lambda_{\rm v,j}+\nonumber\\
    &&(\lambda_{\rm d,i}+\frac{5}{6})\sum_{j=1}^{3}(1+\lambda_{\rm v,j})-(\delta+\frac{5}{2})(1+\lambda_{\rm v,i})\nonumber\\
    &&+\sum_{j\neq i}\frac{[\lambda_{\rm d,j}-\lambda_{\rm d,i}][(1-\lambda_{\rm a,i})^{2}(1+\lambda_{\rm v,i})-(1-\lambda_{\rm a,j})^{2}(1+\lambda_{\rm v,j})]}{(1-\lambda_{\rm a,i})^{2}-(1-\lambda_{\rm a,j})^{2}}\;,\nonumber
\end{eqnarray}
where $\delta=\lambda_{\rm d,1}+\lambda_{\rm d,2}+\lambda_{\rm d,3}$.
The initial conditions for this set of equations are 
\begin{eqnarray}
\label{eq:lambdas}
    \lambda_{\rm a,i}&=&D^\prime_{\rm 1,ini}\lambda_i\;,\\
    \lambda_{\rm v,i}&=&\frac{D^\prime_{\rm 1,ini}\lambda_i}{D^\prime_{\rm 1,ini}\lambda_i-1}\;,\nonumber\\
    \lambda_{\rm d,i}&=&D^\prime_{\rm 1,ini}\lambda_i\;,\nonumber
\end{eqnarray}
where \(\lambda_i\) is the eigenvalue of
\(-S^{(1)}_{i,j}(\vec k,a_0)\), and \(D^\prime_{\rm 1,ini}\) is the normalized growth factor \(D^\prime_1\) at an initial time \(a \sim 10^{-5}\).
At this initial time the Zel'dovich approximation is accurate enough, hence we can compute the 1st order initial displacement field \(S^{(1)}_{i,j}(\vec k,a_0)\) from the realisation of the density contrast.
Then we diagonalise \(S^{(1)}_{i,j}(\vec k,a_0)\) to get its eigenvalues. Finally, by solving Eqs.~(\ref{eq:NGS}) we can get the collapse time for each particle. The collapse time is defined as the moment of collapse of the first axis of the ellipsoid collapsed, \(\lambda_{\rm a,i}\rightarrow 1\).

Collapsed particles may become part of dark matter halos or of the filament network that connects them. Halos are constructed with an algorithm that mimics their hierarchical formation: (i) for each collapsing particle the code checks their 6 neighbours in Lagrangian space; (ii) a particle that collapses before its neighbours becomes a 1-particle halo; (iii) a collapsing particle may accrete on a halo if it "touches" it in Lagrangian space. The accretion condition is satisfied if, after the halo and the particle are moved using Eq.~(\ref{eq:particle}), their distance is less than a threshold distance (discussed below); (v) collapsed particles that don't accrete on halos become filaments, and may accrete later if a neighbour is accreted to a halo; (vi) halo mergers are checked each time a collapsing particle "touches" two halos, the merger takes place if the halos, moved again with Eq.~(\ref{eq:particle}), get nearer than a threshold distance.

In the {\sc pinocchio} code, the threshold distance for the collapsed particles being accreted into halos is defined as
\begin{equation}
d_{\rm thr}^2=\left\{
\begin{array}{lr}
(f_{\rm a} R^e)^2+(f_{200}R)^2,D_1^\prime\sigma\leq \sigma^\prime\;, &\\
\\
\Big\{f_{\rm a} R^e[1+f_{\rm ra}(D_1^\prime\sigma-\sigma^\prime)]\Big\}^2+(f_{200}R)^2,D_1^\prime\sigma> \sigma^\prime\;, &
\end{array}
\right.
\end{equation}
where $R=(M_h)^{1/3}$, and $M_h$ is the halo mass. The threshold distance for merging between halos is defined as
\begin{equation}
d_{\rm thr}^{\prime 2}=\left\{
\begin{array}{lr}
(f_{\rm m} R_{\rm lar}^e)^2+(f_{200}R_{\rm lar})^2,D_1^\prime\sigma\leq \sigma^\prime\;, &\\
\\
\Big\{f_{\rm m} R_{\rm lar}^e[1+f_{\rm rm}(D_1^\prime\sigma-\sigma^\prime)]\Big\}^2+(f_{200}R_{\rm lar})^2,D_1^\prime\sigma > \sigma^\prime\;, &
\end{array}
\right.
\end{equation}
where $R_{\rm lar}=(M_{h,{\rm lar}})^{1/3}$, with $M_{h,{\rm lar}}$ is the mass of the larger halo, $\sigma$ is the variance of the linear density contrast on the grid, and is a function of time.
\(\sigma^\prime \) is a free parameter that controls the change of the "virial radius", and \(\{e, f_{\rm a},f_{\rm m},f_{\rm ra},f_{\rm rm}, f_{200}\} \) are additional free parameters that have been calibrated with N-body simulation in $\Lambda$CDM. Due to the enhancement of gravitational force in G3, the accretion and merging conditions are easier to be satisfied in G3, even for the same halo mass. Via the Poisson equation, one can absorb the gravitational enhancement into the rescaled density and eventually the mass:  
\begin{equation}
 \frac{\partial^{2}\Psi}{a^{2}}=4\pi G_{\rm eff}\delta\rho_{\rm m}=4\pi G\delta\rho_{\rm m,eff}\;,
\end{equation}
where \( \delta\rho_{\rm m,eff}=\mu \delta\rho_{\rm m} \). 
The true halo mass is \( M_h=\bar\rho_{\rm m}\int_0^{r_{\rm vir}}4\pi r^2\delta\rho_{\rm m} dr\). After the redefinition, the effective halo mass reads
\begin{equation}
\label{eq:halomass}
 M_{h,{\rm eff}}=\bar\rho_{\rm m}\int_0^{r_{\rm vir}}4\pi r^2\delta\rho_{\rm m,eff} dr=\bar\rho_{\rm m} \int_0^{r_{\rm vir}}4\pi r^2\mu\delta\rho_{\rm m} dr. 
\end{equation}
Then one can simply replace the effective masses computed with the above formula in the expression for $R$ and $R_{\rm lar}$.
On linear scales, \(\mu^{\rm L} \) is independent of radius. From Eq.~(\ref{eq:halomass}), one can read
\( M_{\rm h,eff}=\mu^{\rm L}(a)M_{\rm h}\). On the nonlinear scale, \(\mu^{\rm NL}(a,\delta) \) also depends on the radius or equivalently on the local density. 
For simplification, we fix \(\delta=200 \), then  \( M_{\rm h,eff}=\mu^{\rm NL}(a,200)M_{\rm h}\).
As a consequence, the 5th force is screened in virialised objects. 
The output halo mass is computed as \( N \times m_{\rm p} \), where \( N \) is the number of particles within the halo and \(m_{\rm p}=4.62\times10^9 M_\odot/h \) is the particle mass.

\section{Results}
\label{sec:results}

We extend the standard {\sc pinocchio} algorithm according to what described in the previous sections. The code, dubbed as ``{\sc g3-pinocchio}'', will soon be available in the official repository. 
In order to compare our results with full N-body simulations, we choose the box size, number of particles and cosmological parameters as in \cite{Barreira:2013eea,Barreira:2014zza}.  
We run ``{\sc g3-pinocchio}'' in a box with size $200~{\rm Mpc}/h$ with \(512^3\) particles. The cosmological parameters we adopt are $\Omega_{\rm m,0}=0.279$, $\Omega_{\Lambda0}=0.721$, $h=0.731$, $\sigma_8=0.997$, $n_s=0.953$, both for G3 and \(\Lambda\)CDM.
In what follows, we describe three different prescriptions for the G3 model, namely ``linearG3'', ``vainG3'' and  ``grG3''. 
For all three G3 models, we compute the displacement fields of dark matter particles with 2LPT as presented in Section 3.
For ``linearG3'', we compute the collapse times with the linear expression for the gravitational slip function, $\mu^{\rm L}(a)$, with no screening. For ``vainG3'', we use the Vainshtein screened gravitational slip function, $\mu^{\rm NL}(a,\delta)$. In ``grG3'' we compute ellipsoidal collapse as in GR: we choose this case as benchmark to demonstrate the effectiveness of the screening mechanism; in fact, the ``grG3'' case can be seen as an extreme case of screening.
A similar approach is taken in \cite{Barreira:2013eea,Barreira:2014zza}, where the authors define a ``Linear model'', equivalent to our ``linearG3'', and a ``Full model'', same as our ``vainG3''. However, their ``QCDM model'' differs from our ``grG3'': the former only modifies the background evolution, while in the latter we modify the 2LPT displacements. As a consequence, our ``grG3''
run already includes the power spectrum enhancement predicted by linear theory, i.e. the amplitude of the power spectrum on large scales is the same in ``grG3'' and ``linearG3'', ``vainG3''.
\begin{figure}
\includegraphics[width=\columnwidth]{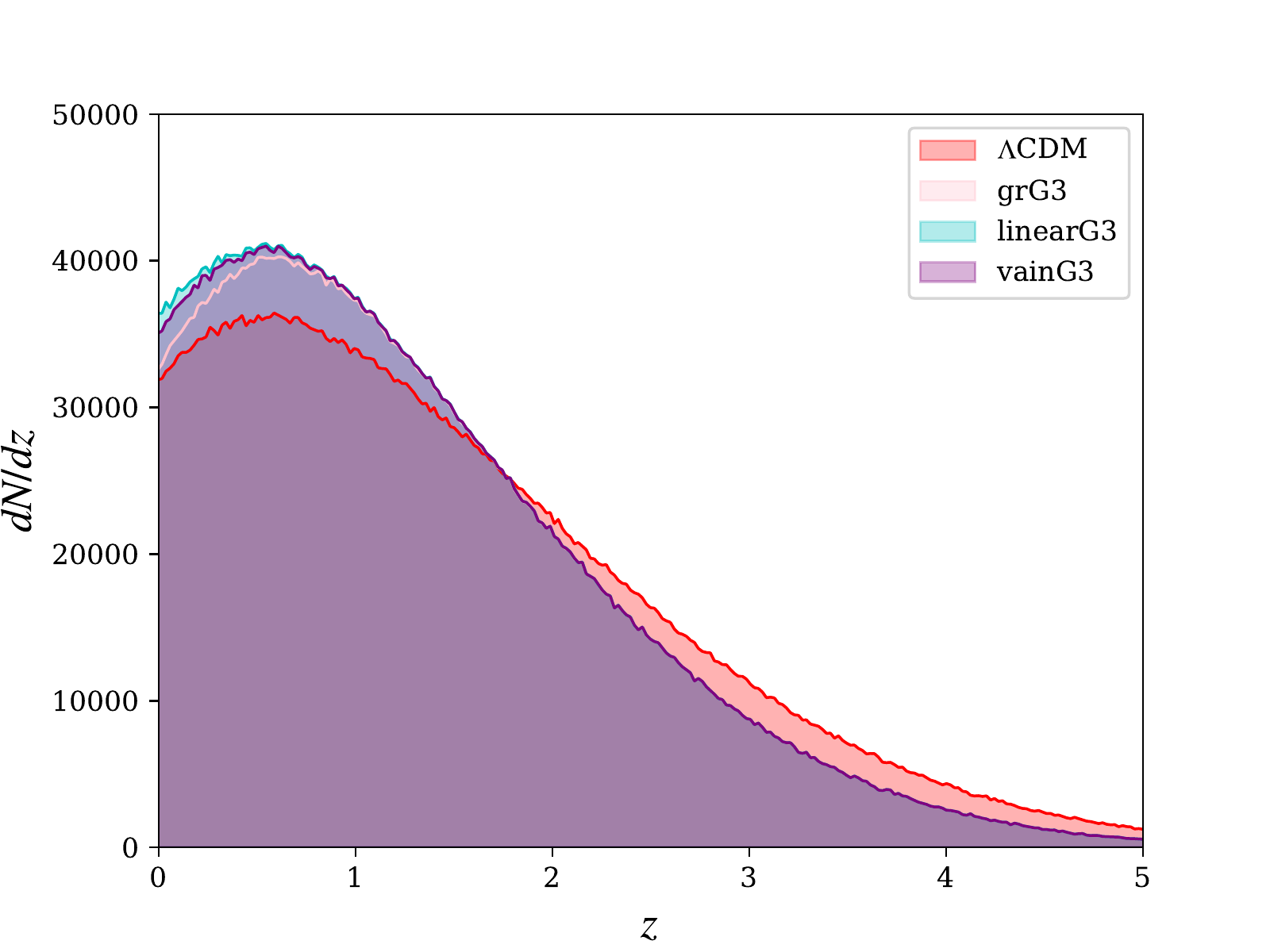}
\includegraphics[width=\columnwidth]{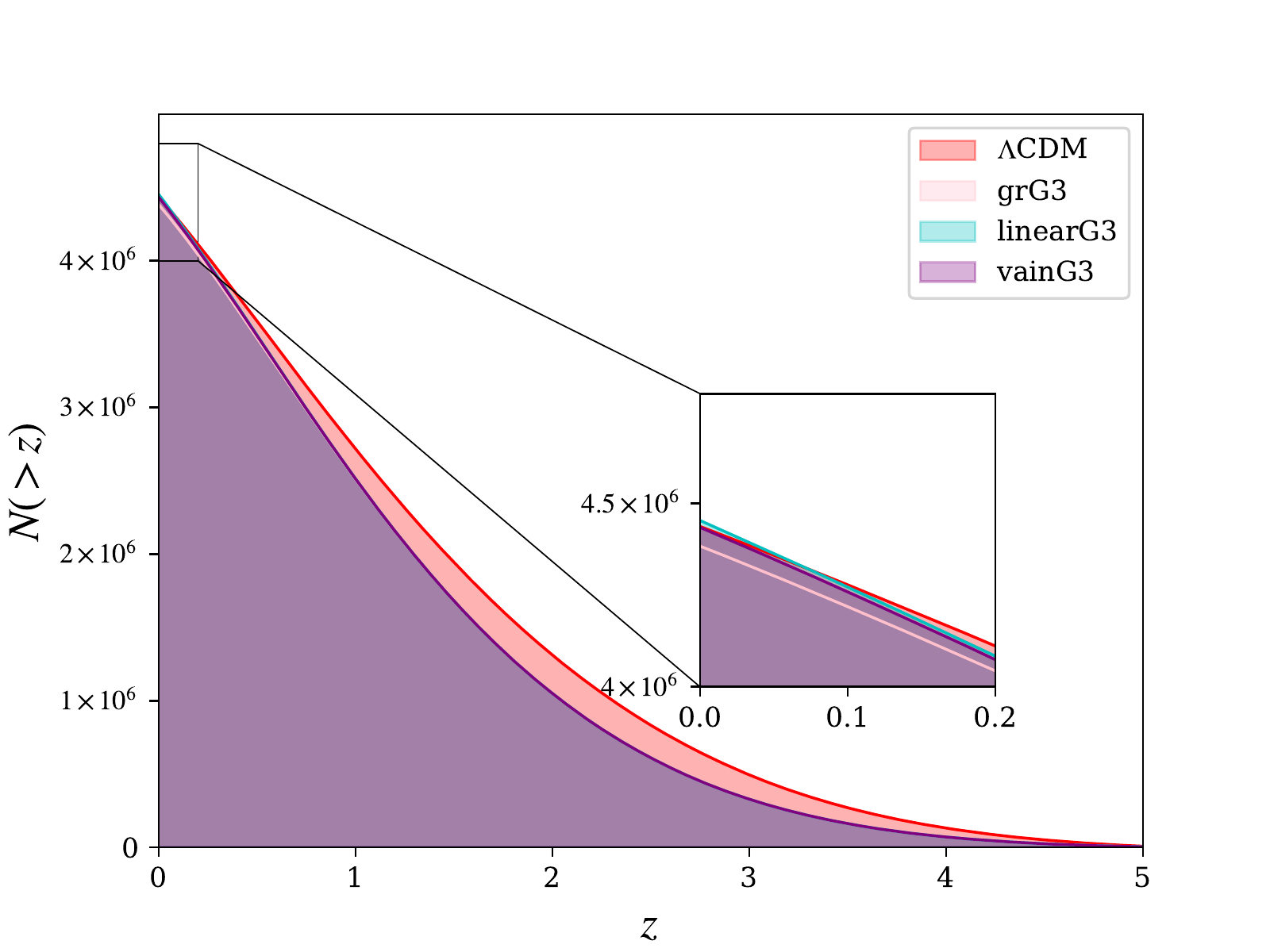}
\caption{Distribution of collapse times for the dark matter particles (top panel) and its cumulative number (bottom panel) for $\Lambda$CDM (red), ``grG3'' (pink), ``linearG3'' (cyan) and ``vainG3'' (purple) as a function of redshift.}
\label{fig:numbers}
\end{figure}

In Fig.~\ref{fig:numbers} we plot the distribution of collapse times for each particle in a smaller simulation with $200^3$ particles and $500$ Mpc/$h$ box size. The upper panel shows the number of collapsed particles for each redshift bin, while the lower panel shows the cumulative numbers.
The red, pink, cyan and purple shaded regions denote respectively $\Lambda$CDM, ``grG3'', ``linearG3'' and ``vainG3'' models.  
From the zoom-in sub-panel, we see that at redshift $z=0$ the cumulative collapsed particle numbers in $\Lambda$CDM and ``vainG3'' are almost the same. This is because we normalize the linear power spectra of both $\Lambda$CDM and G3 models with the same linear perturbation amplitude at $z=0$.  
From the top panel of Fig.~\ref{fig:numbers}, one can see that the $\Lambda$CDM collapsing rate at low redshift is lower than the one of all G3 models. 
On the contrary, the $\Lambda$CDM model has more collapsed particles above redshift $\sim 1.8$. 
This is because \(D^\prime_{\rm 1,ini}\) for \(\Lambda\)CDM is larger than that for G3 by $10-15\%$ (as shown in Fig. \ref{fig:growth}), thus the ellipsoidal collapse for \(\Lambda\)CDM begins with larger density perturbations. This is again an artifact due to the chosen normalization setup. 
In the same figure, the ``vainG3'' collapsing rate is lower than the one of ``linearG3'' between redshifts $0.5$ and $0$. This is due to the $\mu^{\rm NL}$ going back to unity at late times, resulting in a weaker gravitational force with respect to the cases without screening. 
As an extreme screening limit, ``grG3'' further suppresses gravitational collapse. 

The input linear power spectrum can be computed with an Einstein-Boltzmann solver such as CAMB/CLASS \citep{Lewis:1999bs,2011arXiv1104.2932L} or using the analytic fitting formula of \cite{Eisenstein:1997ik}. 
To suppress sample variance in the power spectrum, the moduli of Fourier-space modes of the linear density field are fixed to their expectation value given by the power spectrum, while phases are randomly distributed between $0$ and $2\pi$.
Since the linear growth rate in G3 is larger than the one of $\Lambda$CDM, to reach the same final amplitude the G3 model has to start from a more uniform density initial condition. For this reason, the high redshift matter power spectra in G3 are lower than those in $\Lambda$CDM cases. 
\begin{figure}
	\includegraphics[width=\columnwidth]{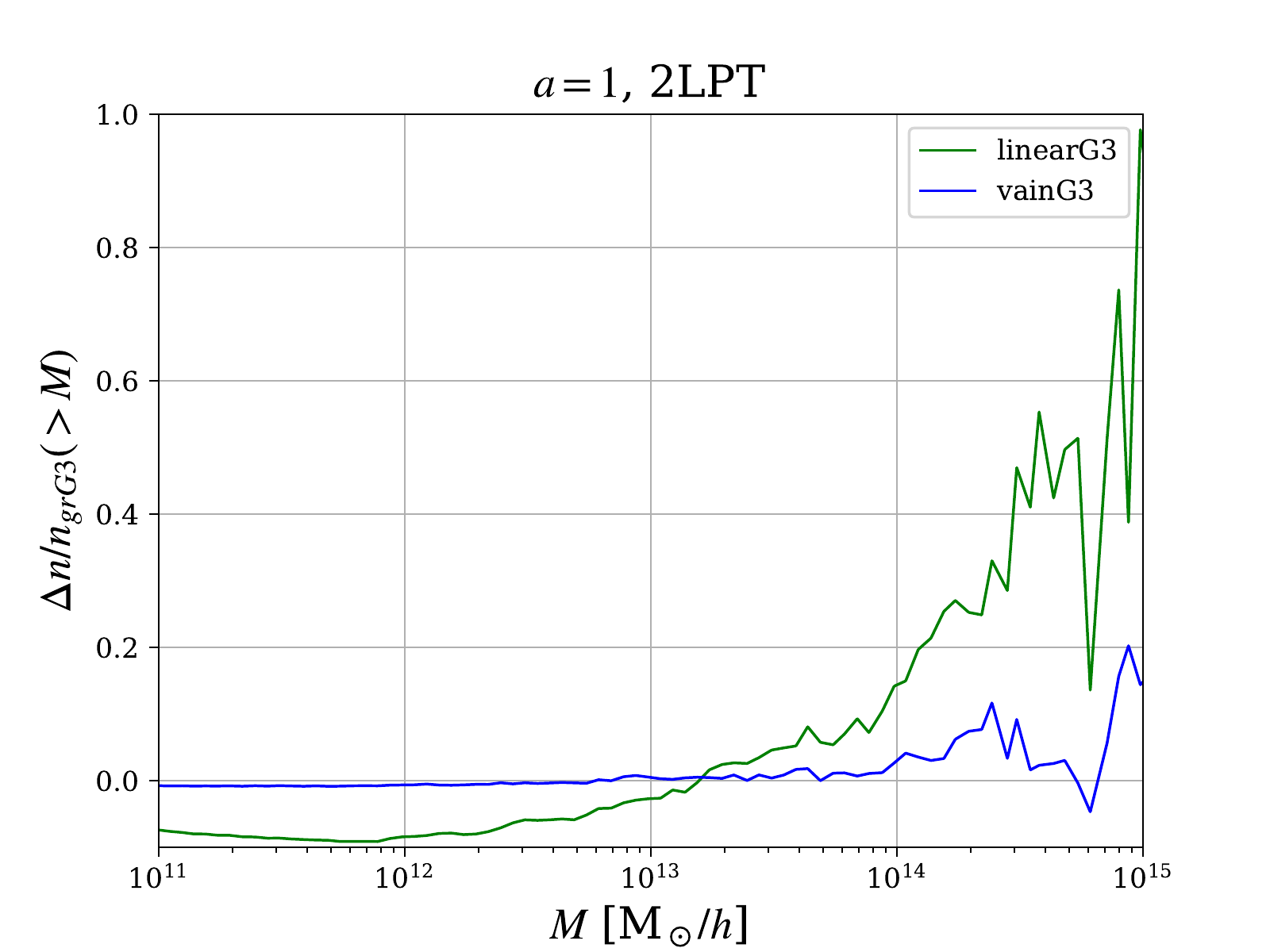}
	\includegraphics[width=\columnwidth]{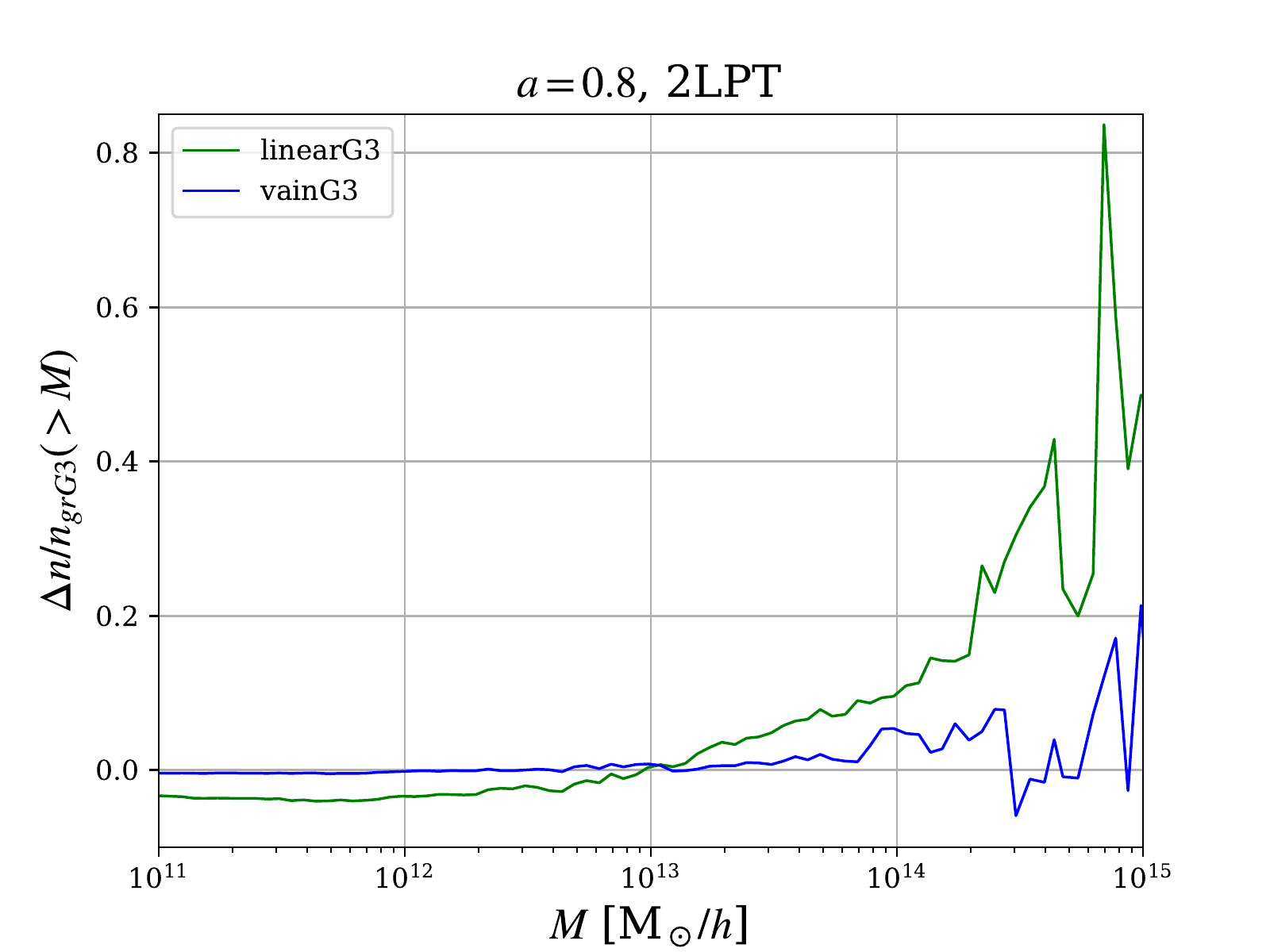}
    \caption{Relative differences in the cumulative mass function with respect to the ``grG3'' case at $a=1$ (top panel) and $a=0.8$ (bottom panel) for our two approaches to ellipsoidal collapse, ``linearG3'' and ``vainG3'', shown with green and blue lines respectively. }
    \label{fig:massfunc1}
\end{figure}

In Fig. \ref{fig:massfunc1} we show the relative differences in the cumulative mass function of ``linearG3'' (green lines) and ``vainG3'' (blue lines) with respect to our benchmark, ``grG3''. The role of the Vainshtein mechanism can clearly be seen in the high mass end, where the fifth force is screened
and the ``vainG3'' case shows a smaller deviation from ``grG3'' than ``linearG3''.

Additionally, we compute the linear halo bias. We use the \texttt{PowerI4} package \footnote{https://github.com/sefusatti/PowerI4} to read the halo catalogues and compute the halo power spectra at different redshifts, then we compute the square root of the ratio between the halo power spectrum and the linear matter power spectrum for \( k<0.2\), namely the linear halo bias.
\begin{figure}
	\includegraphics[width=\columnwidth]{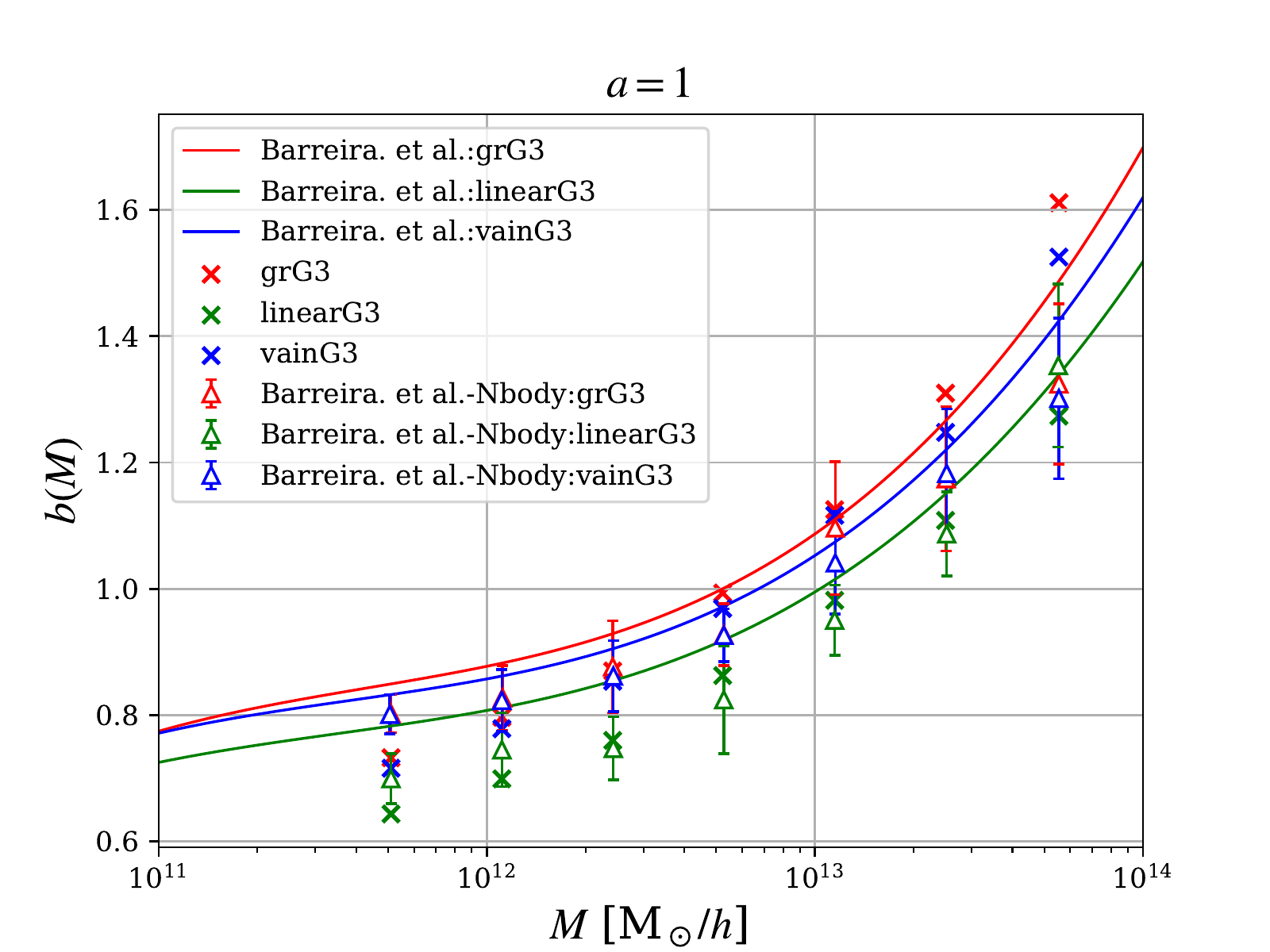} 
	\includegraphics[width=\columnwidth]{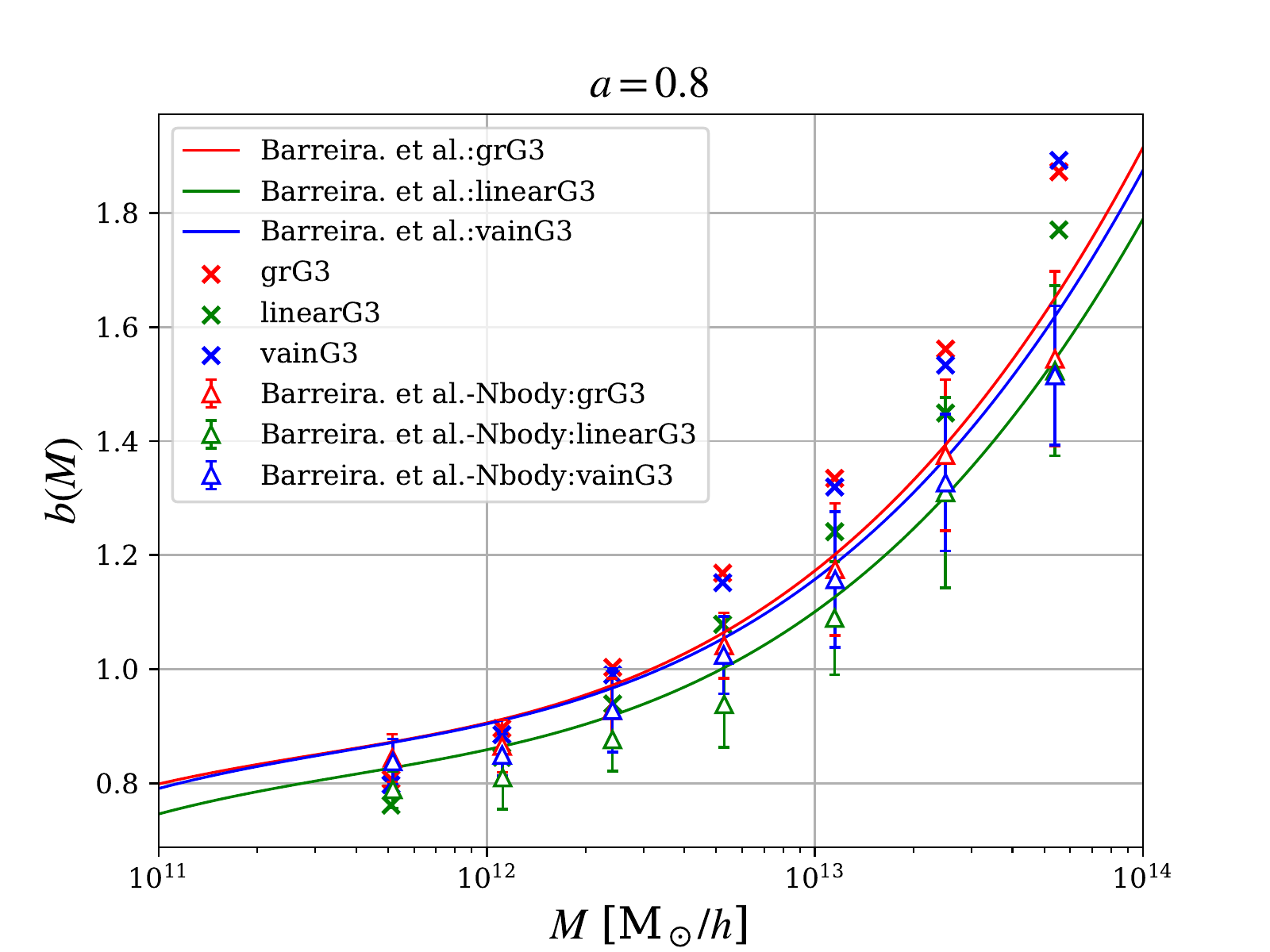}
    \caption{Linear halo bias at $a=1$ (top panel) and $a=0.8$ (bottom panel), as a function of the halo mass. We mark our predictions with cross symbols, and compare to the results of \protect\cite{Barreira:2014zza} showing their simulations results with triangles with errorbars, and their prediction based on the Sheth-Tormen formula with solid lines. We show the results for the ``grG3'' case, the ``linearG3'' case and the ``vainG3'' case with red, green and blue lines and symbols respectively.}
    \label{fig:linearBias}
\end{figure}

In Fig. \ref{fig:linearBias} we compare our results for the halo bias with the N-body simulations of \cite{Barreira:2014zza}. The cross symbols are our results, the solid curves are the best-fitting Sheth-Tormen formalism to the cumulative halo mass function data from simulations, and the triangles with error bars are the measurement of halo bias directly from the simulations. Different colors mark the different approximations for collapse times, as detailed in the legend. It can be seen that our results, despite being computed with a completely different method, match remarkably well the general trend of the N-body simulations.

\begin{figure}
	\includegraphics[width=\columnwidth]{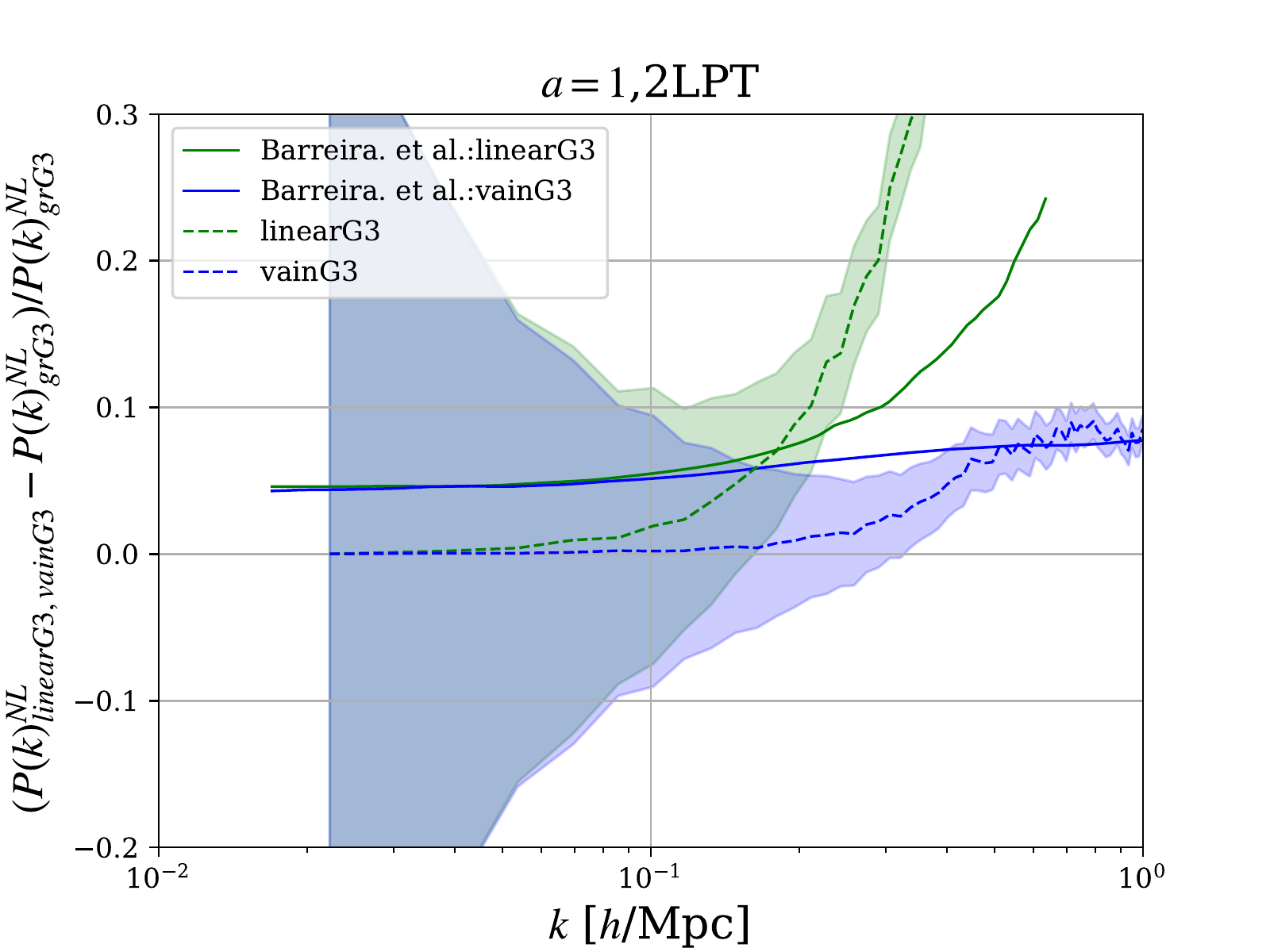} 
	\includegraphics[width=\columnwidth]{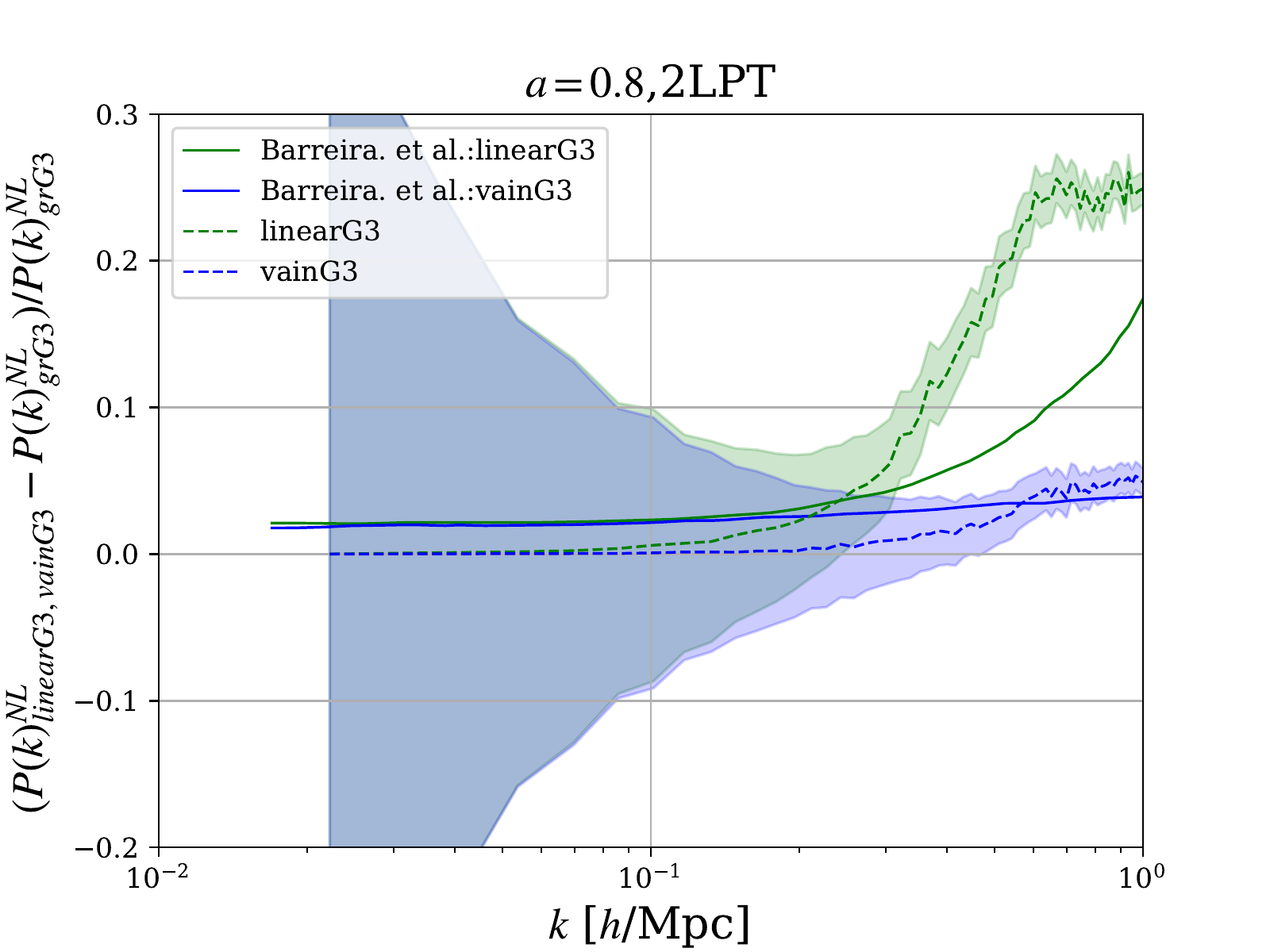}
    \caption{Matter power spectrum relative differences at $a=1$ (top panel) and $a=0.8$ (bottom panel). The green lines mark the ratio of the ``linearG3'' simulations to the ``grG3'' case, while we show the ``vainG3'' case in blue. Our results, measured from {\sc pinocchio} snapshots, are plotted as dashed curves. The shadowed areas mark the 1$\sigma$ errors on the measurements. We also show the results from 
     \protect\cite{Barreira:2013eea} with solid lines, remarking that these ratios are obtained to their QCDM simulations, which does not correspond to our ``grG3'' case exactly since it only includes modifications to the background evolution.}
    \label{fig:mpk}
\end{figure}

Finally, we plot our result for the matter power spectrum in Fig. \ref{fig:mpk}. 
For {\sc pinocchio}, this is obtained by displacing with 2LPT particles that do not belong to halos, while halo particles are distributed around the halo center of mass assuming an NFW profile \citep{Navarro:1996gj,Navarro:1995iw}. 
These power spectra are computed with the \texttt{Pylians} package \footnote{https://pylians3.readthedocs.io/en/master/}. 
For comparison, we also show the simulation results with solid curves. The relative differences are with respect to the ``grG3'' model in our case, and with respect to the ``QCDM'' model in the case of the simulations. One can see that, at large scales, the two results are discrepant: in particular, the \cite{Barreira:2013eea} ratios also show the amplitude difference in the power spectra that is due to the different linear evolution between G3 and $\Lambda$CDM, while ours do not. As mentioned previously, this is due to the fact that our ``grG3'' is not exactly the same as ``QCDM'', since we modify both the background and the Lagrangian perturbations up to the 2nd order while the ``QCDM'' model in \cite{Barreira:2013eea} only modifies the background evolution.  For this reason, we cannot do a quantitative comparison with \cite{Barreira:2013eea}, although it can be seen that our results reproduce the general trend of the simulations.
We report in Appendix \ref{app:catalog} a more detailed analysis of our halo catalogues. We compare the halo properties of matched and unmatched catalogues, including the halo mass distributions, the halo mass functions and halo power spectra.
\section{Conclusions}
\label{sec:con}
This work presents 1st and 2nd order Lagrangian perturbation theory for the cubic Galileon model, as well as its ellipsoidal collapse dynamics.  
As in the case of $\Lambda$CDM, both the 1st and 2nd order growth factors of the cubic Galileon model are scale independent, since the extra scalar field which drives the accelerated expansion is effectively massless. Its Compton wavelength is indeed on the Hubble horizon scale: below this scale, the modifications of gravity keep constant in space and vary slowly with time. Since both growth functions are larger that their $\Lambda$CDM counterparts, as shown in Fig.~\ref{fig:growth}, clustering in enhanced.
In particular, the maximum enhancement of gravity happens at redshift zero, when the effective gravitational constant  is about twice the Newton constant. 
Once we include the Vainshtein screening mechanism in the high local curvature regime, the gravitational interaction restores to the standard GR case. 

Using both the extension of standard 2LPT and ellipsoidal collapse to the G3 model, we create a new branch of the {\sc pinocchio} code, dubbed as {\sc g3-pinocchio}: our extension provides a fast tool to generate approximated dark matter halo catalogues with the G3 modified gravity model. 
We run the {\sc g3-pinocchio} code in a box with size $200~{\rm Mpc}/h$ and \(512^3\) particles and study the properties of the obtained halo catalogues at different redshifts. 
To illustrate the effect of Vainshtein screening, we run three different implementations of ellipsoidal collapse, namely ``linearG3'', ``vainG3'' and ``grG3''. From these realisations we compute the cumulative mass function, the linear halo bias and the matter power spectrum. 
We find that, as for other types of modified gravity, the Vainshtein screening mechanism in cubic Galileon also suppresses the extra gravitational force in the high mass end of the mass function. 
We compute the linear halo bias and the matter power spectrum and perform a qualitative comparison to N-body simulations, showing that we can reproduce all trends remarkably well.  
Given the significantly reduced computational time required to generate halo catalogues with {\sc pinocchio} with respect to more computationally expensive N-body simulations, our implementation provides an optimal tool for the fast generation of large sets of realisations. Additionally, the code can be readily extended to include other MG models that feature Vainshtein screening.

\section*{Acknowledgements}
We thank Noemi Frusciante, Simone Peirone, Alessandra Silvestri, Jianhua He and Cheng-Zong Ruan for various discussions. We are grateful to the anonymous referee for useful comments which lead us to improve our results. YLS and BH are supported by the National Natural Science Foundation of China Grants No. 11973016. CM acknowledges support from a UK Research and Innovation Future Leaders Fellowship (MR/S016066/1).

\section*{Data Availability}

The data underlying this article will be shared on reasonable request to the corresponding author.



\bibliographystyle{mnras}
\bibliography{example} 




\appendix

\section{Halo catalogues}
\label{app:catalog}
\begin{figure}
	\includegraphics[width=\columnwidth]{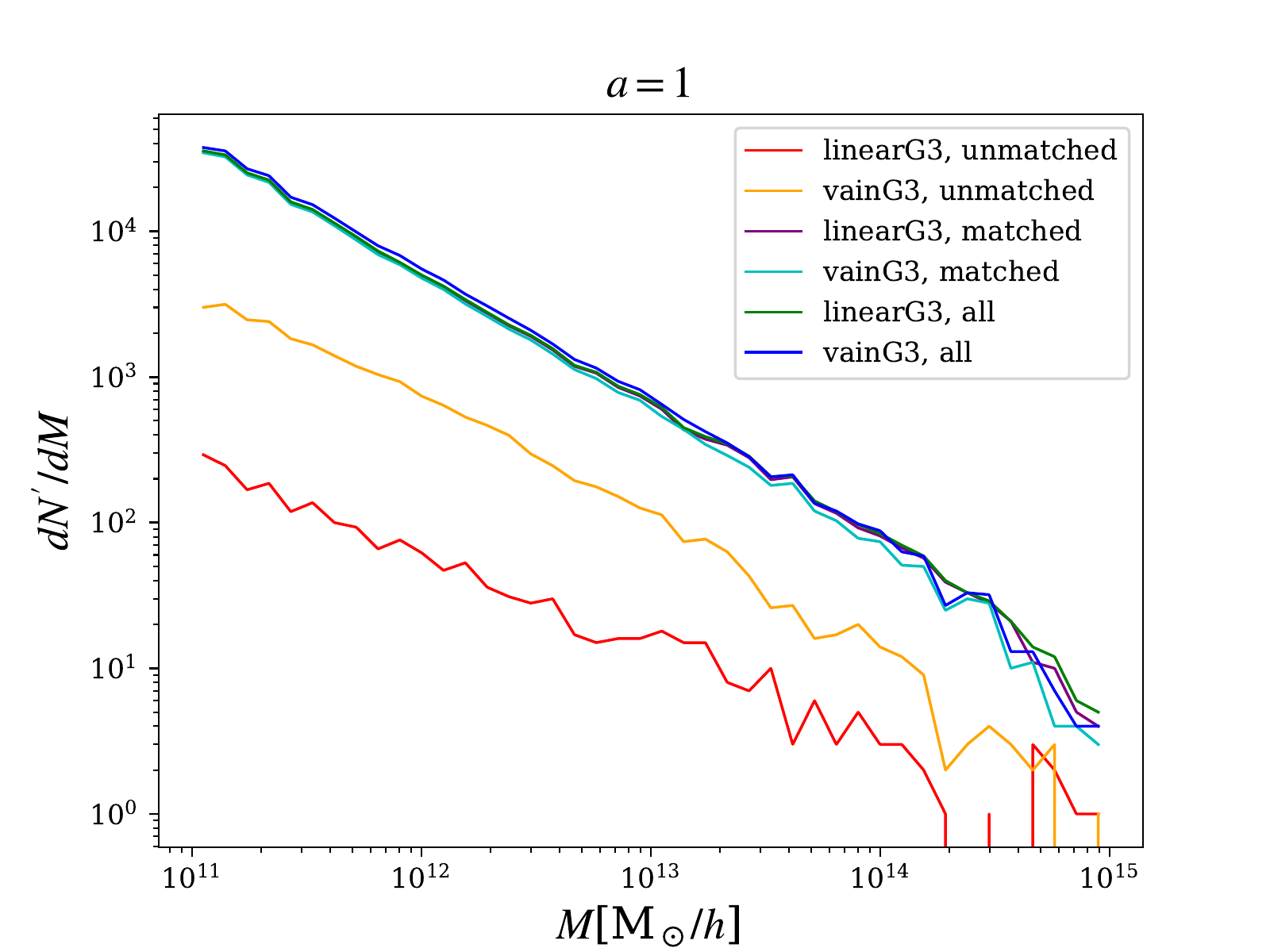}
	\includegraphics[width=\columnwidth]{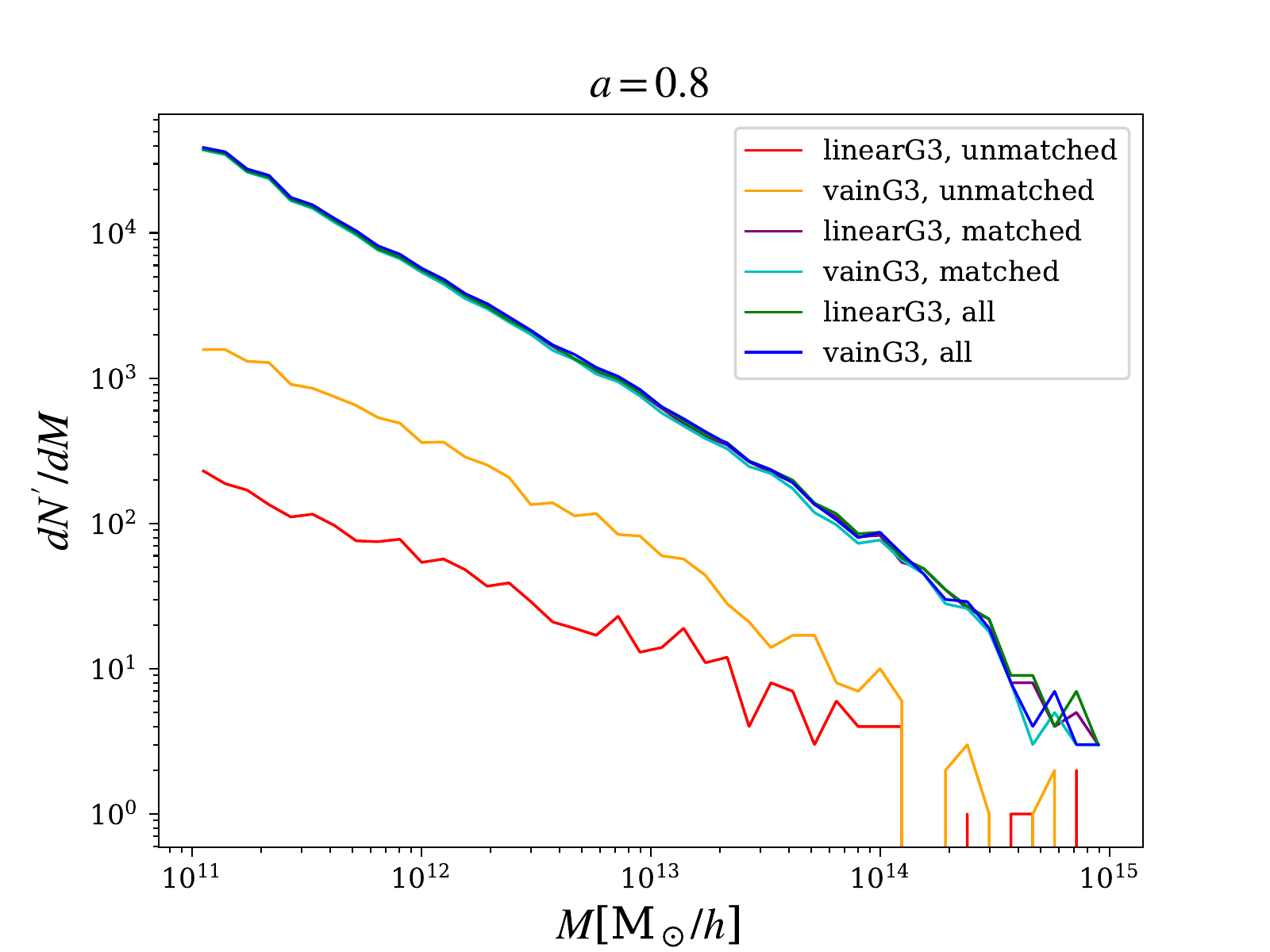}
    \caption{Halo number density $dN/dM$ as a function of the halo mass at $a=1$ (top panel) and $a=0.8$ (bottom panel). We show results for the `linearG3'' and `vainG3'' cases, for the whole halo catalogue(marked as {\it all}, shown with green and blue lines) as well as for the catalogue split in matched (purple and cyan lines) and unmatched (red and orange lines) halos.}
    \label{fig:numHalo}
\end{figure}

We describe here a more detailed analysis of the halo catalogues.  
Since we use the same initial conditions for the different implementations of the modified gravity model, namely 
``linearG3'' and ``vainG3'', we can match the IDs of halos that are formed in both simulations.  We call them ``matched'' halos, while halos that are only present in one of the two realisations are called ``unmatched''.
In Fig. \ref{fig:numHalo} we show the halo number density for the matched and unmatched halos. 
One can see that most halos appear in both realisations, while only $\sim 1\%$ of the halos in ``linearG3'' are unmatched and $\sim 10\%$ halos in the ``vainG3'' are unmatched.  
\begin{figure}
	\includegraphics[width=\columnwidth]{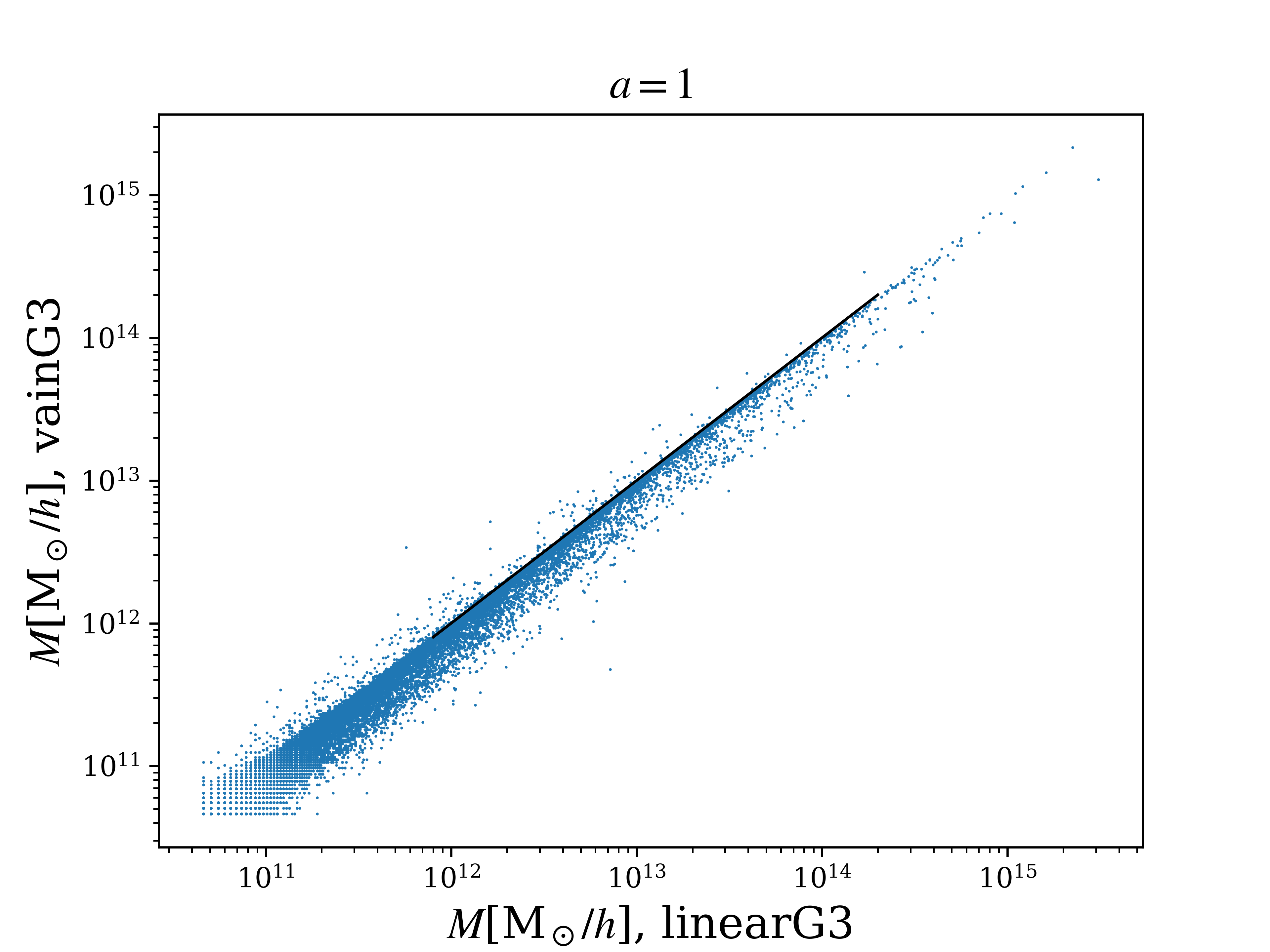} 
	\includegraphics[width=\columnwidth]{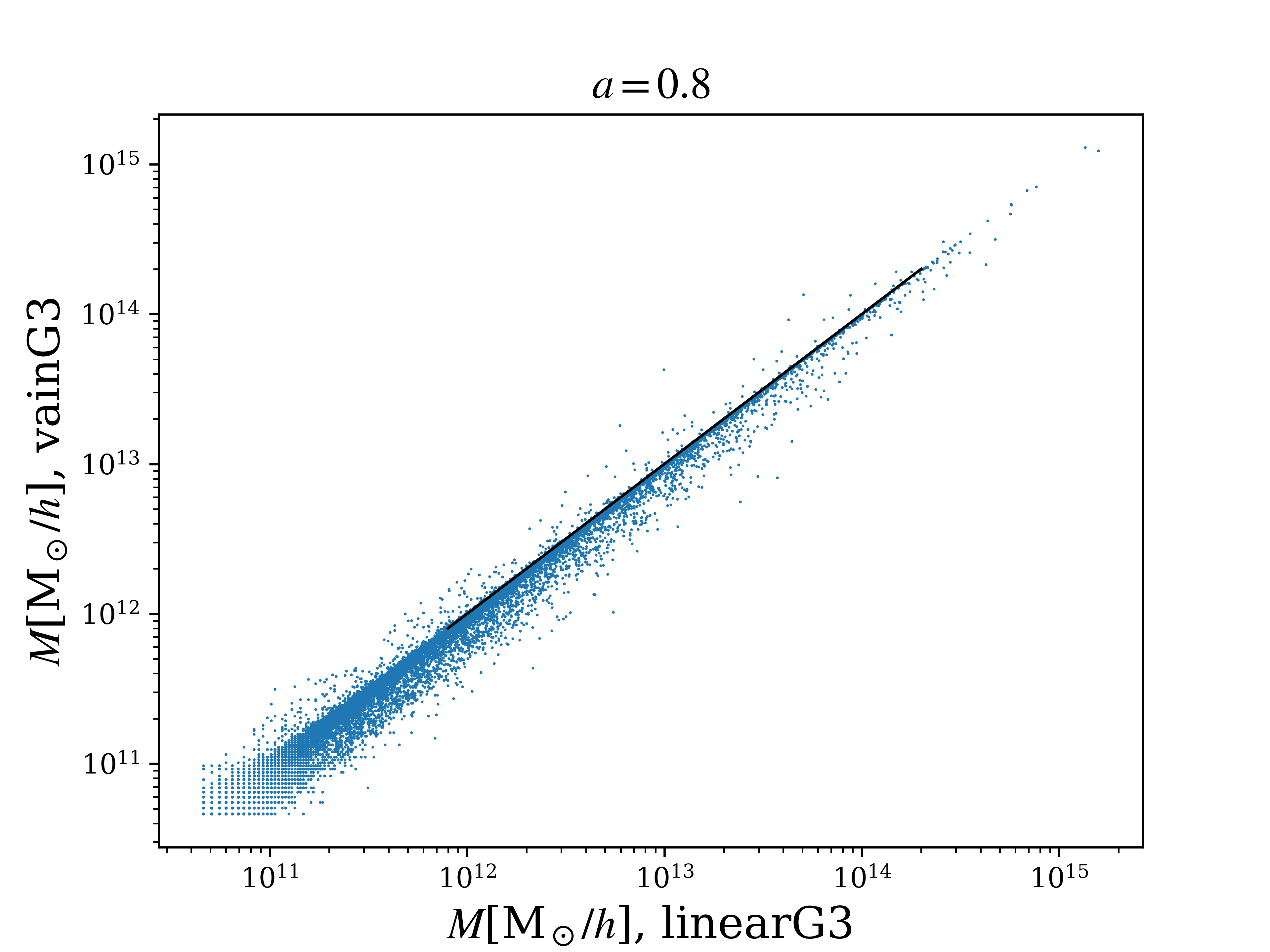}
    \caption{Mass distribution of the matched halos at $a=1$ (top panel) and $a=0.8$ (bottom panel) for the ``linearG3'' versus ``vainG3'' cases.}
    \label{fig:massdis}
\end{figure}
In Fig. \ref{fig:massdis} we show the mass distribution of the matched halos for the ``linearG3'' and ``vainG3'' prescriptions. Because the modification to the gravitational slip is stronger in the linear case with respect to the nonlinear ``vainG3'' one, the halo masses shift to the massive direction in the ``linearG3'' case (i.e. they mostly sit below the black solid line). 
\begin{figure}
	\includegraphics[width=\columnwidth]{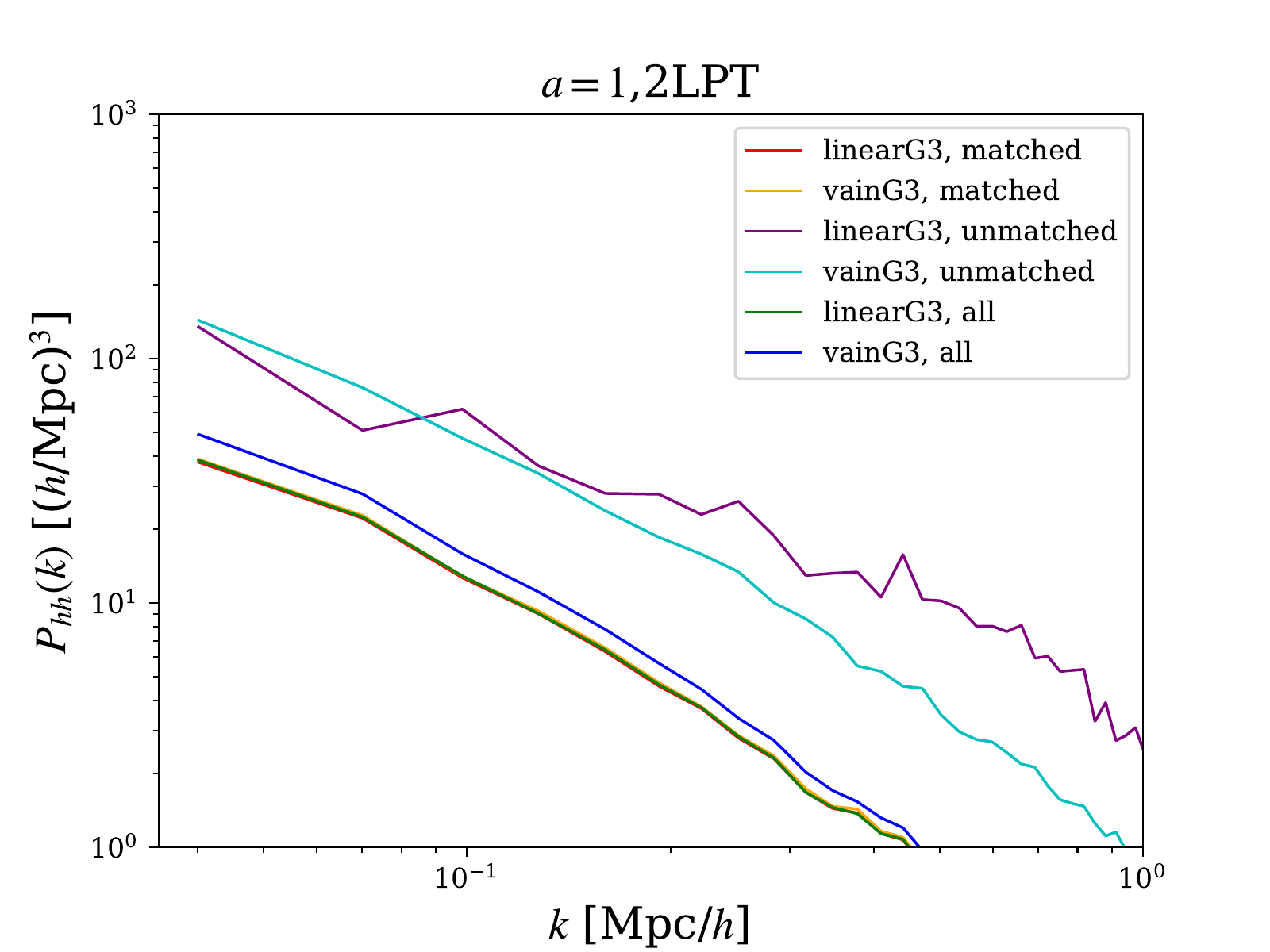} 
	\includegraphics[width=\columnwidth]{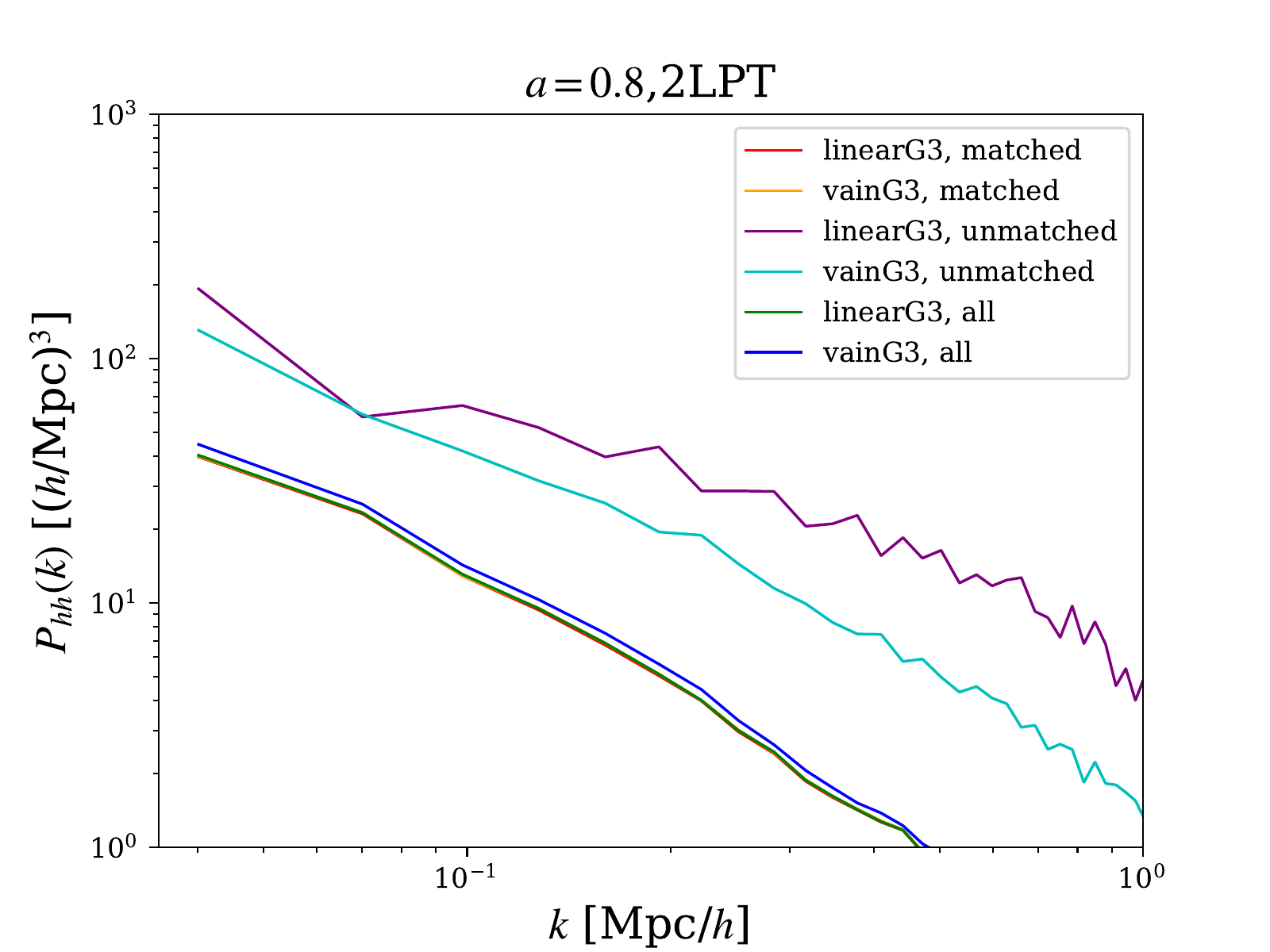}
    \caption{Halo power spectra at $a=1$ (top panel) and $a=0.8$. As above, we show results for the whole catalog, the matched and unmatched halos.}
    \label{fig:Phh}
\end{figure}
In Fig. \ref{fig:Phh} we show the halo power spectra computed for the whole catalog, and for the catalogue split in matched and unmatched halos. The halo power spectra are computed with \texttt{PowerI4}, considering all halos with at least 100 particles. It can clearly be seen that the halo power spectra of the matched halos are very similar for both ``linearG3'' and `vainG3'', while they are significantly different for the unmatched halos. One can conclude that the differences in the final halo power spectra actually come from the unmatched halos.   
\begin{figure}
	\includegraphics[width=\columnwidth]{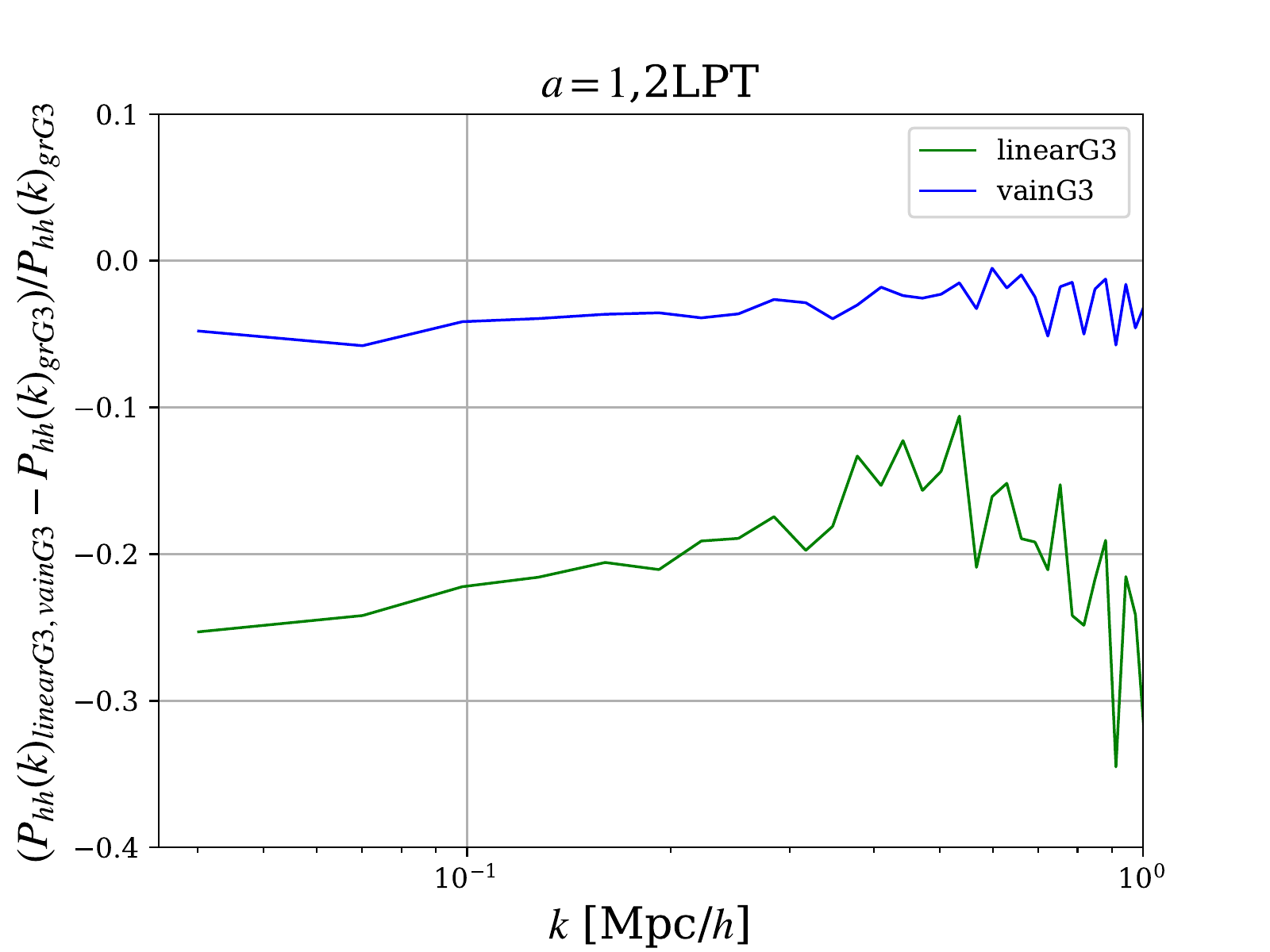} 
	\includegraphics[width=\columnwidth]{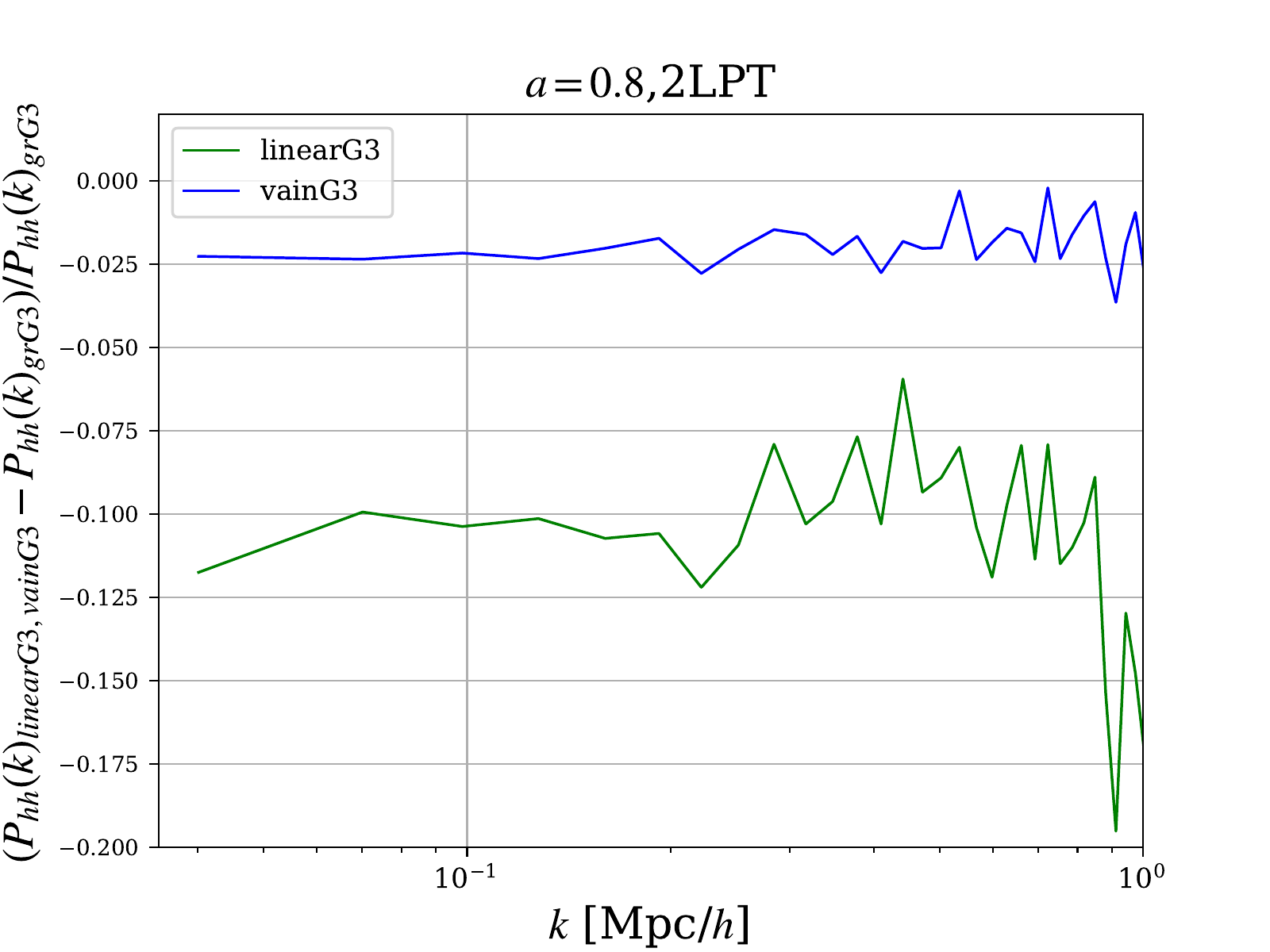}
    \caption{Relative differences in the halo power spectra at $a=1$ and $a=0.8$, shown for the ``linearG3'' prescription (green line) and the ``vainG3'' prescription (blue line). In both cases the ratio is computed with respect to the ``grG3'' case.} 
    \label{fig:PhhRes}
\end{figure}
Finally, in Fig. \ref{fig:PhhRes} we show the relative differences in the halo power spectra with respect to the ``grG3'' model. At $a=1$ we find a difference of $\sim 20\%$ on all scales for ``linearG3''. Such difference is strongly reduced for ``vainG3'', highlighting once again the effectiveness of the Vainshtein mechanism in screening the MG fifth force. Similar results can also be seen for $a=0.8$, with differences of $\sim 10\%$ and $\sim 2\%$ respectively for ``linearG3'' and ``vainG3''.


\bsp	
\label{lastpage}
\end{document}